\documentclass{elsarticle}
\usepackage{float}
\usepackage{bm}
\usepackage{xcolor}
\usepackage{amsmath}
\usepackage{amssymb}
\usepackage{graphicx}
\usepackage{lineno}

\journal{Chaos, Solitons \& Fractals}

\begin{document}
%\begin{linenumbers}
\begin{frontmatter}

\title{External charged debris in a flowing plasma : charge fluctuation induced
complexity}
\author{Bikramjit Joardar, Hitendra Sarkar, and Madhurjya P.\ Bora\corref{cor1}}

\address{Physics Department, Gauhati University, Guwahati, 781014, Assam, India}
\cortext[cor1]{Corresponding author: mpbora@gauhati.ac.in}
\begin{abstract}
In this work, we investigate the response of a flowing $e$-$i$ plasma
to embedded external charged debris, focusing on the periodic debris
charge fluctuations that can trigger complex phenomena such as chaos
and nonlinear Landau damping. We employ both kinetic and fluid simulations
to analyse the plasma response to the time-dependent debris charge.
Our findings indicate that the nature of nonlinear response can be
considerably different for fluctuating positively charged external
debris from a negatively charged debris. The simulations show that
the debris charge fluctuation causes damping of the ion-acoustic wave
as the debris velocity nears the ion-acoustic speed through nonlinear
Landau damping and wave-wave interactions. We also present a theoretical
framework to support the simulation findings. Our findings provide
critical insights into debris-plasma interactions which may be useful
in applications involving space debris management.
\end{abstract}

\begin{highlights}
\item {\bf Model}: Investigates the effect of periodic charge fluctuation of external debris in a flowing $e$-$i$ plasma.

\item {\bf Nonlinear complexity}: Demonstrates emergence of debris charge-dependent chaos and nonlinear Landau damping (NLLD) using kinetic and fluid simulations.

\item {\bf Nonlinear modeling}: Theoretical analysis confirms simulation findings.

\item {\bf New insights}: Positively charged debris induces stronger localised chaos compared to negatively charged debris, with latter exhibiting NLLD.

\item {\bf Potential application}: Findings offer new insights into the plasma-debris interactions which may add another dimension to the debris detection scenario in low Earth orbit plasma.

\end{highlights}

%% Keywords
\begin{keyword}
Plasma-Debris Interaction \sep Charge Fluctuation \sep  Chaos
\sep Nonlinear Landau damping
%% keywords here, in the form: keyword \sep keyword

%% PACS codes here, in the form: \PACS code \sep code

%% MSC codes here, in the form: \MSC code \sep code
%% or \MSC[2008] code \sep code (2000 is the default)

\end{keyword}

\end{frontmatter}

\section{Introduction}

Impurities or external entities in a typical electron-ion plasma is
not a new subject and the arena of scientific research involving such
``foreign'' entities in plasma environments is dotted with numerous
studies, which essentially began with Irving Langmuir's pioneering
work on plasma sheath in the 1920s and 1930s \citep{irving1,irving2,irving3,irving4,irving5,irving6}.
One of the most distinctive branches which stemmed out of such plasma
physics research is the so-called ``dusty plasma'', which has become
one of the core research areas in plasma physics \citep{whipple,goertz,rao,shukla}.
Another such area, which came to the notice of the scientific community
is the interaction of the plasma with charged external or foreign
bodies, when the importance of scientific studies of such interactions
was realised during the early days of space exploration, in the 1960s,
when several space missions began studying how satellite and spacecraft
interact with space plasma environments \citep{nasa1,alan,samir,gian}.
In recent years, research in this area has really accelerated, especially
after the subject of space debris in relation to the safety of the
satellites and space stations in low Earth orbits (LEOs) has been
recognised as an important research area. Usually, space debris at
LEO are detected with satellite and ground-based sensors operating
at optical and radio ranges and these methods are not effective in
detecting smaller debris (size $\lesssim1\,{\rm cm}$) \citep{size1},
though NASA's Haystack Ultrawideband Satellite Imaging Radar (HUSIR)
has been able to provide a map of LEO debris down to $\sim5.5\,{\rm mm}$
\citep{size2}. Recent techniques, based on detection of plasma oscillations
excited by these smaller debris, have been tested successfully to
detect some of these debris \citep{AMOS}. In essence, the number
of proven and effective methods to detect these smaller debris, is
by no means complete and we continue to look toward understanding
different physical manifestations of plasma-debris interaction. And
this work is an effort to understand the intricacies of these interactions
by allowing more realistic physical conditions.

In recent times, our understanding about plasma-debris interaction
has been steadily progressing due to several theoretical, experimental,
and numerical studies \citep{sen2015nonlinear,jaiswal2016experimental,arora2019effect,sarkar2023response,sarkar2025exploring,mrid}.
It is now definitively proved both theoretically and through numerical
simulations that depending on the nature of the charged debris,
different nonlinear waves in the ion-acoustic regime can get excited
by moving debris \citep{sarkar2023response,mrid}. While we usually
see dispersive shock waves due to a positively charged external debris,
pinned solitons are formed due to phase space trapping through ion-ion
counter streaming instability (IICSI) {for a} negatively charged
debris \citep{mrid}. Some very recent laboratory and numerical simulation
studies have looked into the plasma-debris interaction in multi-ion
plasma \citep{sarkar2025exploring}. Several other studies have focused
on shape, dimension, and speed of the debris on the propagation of
these solitons theoretically and with experiments \citep{truitt2020a,arora2019effect},
as well as bending and acceleration\textbf{ }\citep{acharya2021bending},
damping (both collisional and Landau) \citep{truitt2020b}\textbf{
}of the excited structures.

In this work, we plan to explore a yet unexplored area related to
plasma-debris interaction by allowing the charge in the external debris
to vary over time along with the presence of a stochastic component.
There are good reasons as to why the charge on external debris should
vary. Some seemingly plausible causes are periodic photoemission due
to eclipse transitions (especially debris in the LEO), passage of
the debris through the regions of density and/or temperature gradient
of the ambient plasma, plasma wave interactions, intrinsic forcing
due to spin or tumbling motion of the debris, and external modulations
viz. during spacecraft remediation processes, among others. Out of
these scenarios, the most important one is perhaps the debris charge
modulation due to plasma wave interactions as this is an inherent
process and possibly cannot be avoided. Debris moving at or near ion-acoustic
speed is expected to make the charge on debris oscillate in time at
ion-plasma frequency $(\omega_{pi})$, through ion-acoustic wake oscillations
\citep{gurudas}. As an ion-acoustic wave (IAW) passes over the debris,
it periodically alters the ion and electron densities. Since the ion
and electron currents depend sensitively on these densities, the wave
creates an imbalance in the ion and electron fluxes to the debris
surface. This imbalance, in turn, disturbs the equilibrium floating
potential of the debris and the debris charge adjusts dynamically
to restore the balance, leading to a periodically oscillating debris
charge. So, basically IAWs excited by the external charged debris,
itself can cause the periodic fluctuation of the debris charge at
frequencies $\sim\omega_{pi}$, which is very much inherent to the
debris-plasma interaction scenario. This process is very much similar
to the process of dust-charge fluctuation in a dusty plasma \citep{gurudas,jana}.
This inherent charge fluctuation of the debris can be thought to be
a forcing term, which can result in parametric coupling and can give
rise to a wide range of nonlinear phenomena including chaos, sheath
instability, nonlinear Landau damping \citep{nlandau} (NLLD) and
can also affect the local turbulence. In this analysis, we primarily
focus on two effects -- chaos and NLLD, the signatures of which are
most prominent. While detecting chaos is pretty straight forward,
the same is not true for NLLD, which requires more than one confirmation
methods. We believe that the periodic oscillation near $\omega_{pi}$
can trigger NLLD, which can then limit the wave amplitude, providing
a feedback mechanism on the charging process and may eventually affect
the nonlinear structures \citep{nlandau1}. {On the other hand, as
chaotic oscillations can imply the breakdown of coherent wave-particle
energy exchange, it can act as a probe for the
wave-particle interaction limit \citep{torus1,chaoslimit}. We carry out our 
primary investigation with  a 1D electrostatic particle-in-cell (PIC) method \citep{birdsall}.
We shall also use the results of a flux-corrected-transport (FCT) method \citep{boris1973} to support our findings.

\subsection{Why 1D?}
While the real physical system is invariably 3D, one can get valuable physics information from 1D models very easily. There are three pressing reasons why in many cases, we resort to a 1D modeling. First, the symmetry. As there is no preferential direction, any direction is equivalent. Second, in many cases, the dominant dynamics happen only in one direction viz.\ direction of the flow. So, one can analyse the essential physics from studying the system in that dominant direction only. The third and the most important is the ease of calculation in 1D models as compared to 2D or 3D. In 2D or 3D, the resolution required to resolve Debye-scale fluctuations in the vicinity of the external debris will be prohibitively large.

In our case, as we shall see, the major dynamical activities are along the longitudinal direction or the direction of plasma (or debris) flow. So as long as we do not expect the transverse activities to affect the dynamics considerably (which is in our case), we can safely consider the 1D model. By assuming a slab-symmetric or axisymmetric configuration and a uniform plasma flow, the system effectively reduces to one dimension without significant loss of physical generality. Note that when an extended object (like a block or slab) move in a plasma, majority of the dynamics happen across the plasma-facing side of the debris, which is essentially like a 1D situation as all interactions across the plasma-facing side should be same. The situation would have been completely different had there been a magnetic field, which would have necessarily required transverse dynamics, thus forcing a multi-dimensional model. 

Moreover, the 1D model allows us to use tractable mathematical techniques for analysing the chaos such as Lyapunov exponent analysis, Kolmogorov--Arnold--Moser (KAM) theory, and finite time Lyapunov exponent (FTLE) maps, which would be analytically and computationally prohibitive in higher dimensions. Hence, the 1D approach is not only physically motivated but also essential for uncovering the fundamental mechanisms governing plasma-debris interactions, which can be later benchmarked in detail through 2D or 3D models (which is beyond the scope of this work). The 1D model thereby serves as a minimal yet powerful framework to isolate and study these nonlinear mechanisms.}

\subsection{ Organisation of the paper}

The paper is organized as follows. In Section II, we present the results
of kinetic and fluid simulations with the help of our \emph{h}-PIC-MCC
and \emph{m}FCT codes, respectively. In Section III, we provide a
theoretical analysis of the results obtained in Section II. We provide
an additional analysis of the results obtained in earlier sections,
in terms of power spectrum analysis in Section IV. In Section V, we
conclude. We also provide a brief analysis of the effect of fluctuation
of debris charge on turbulent dynamics. { In the Appendix, we discuss the validity of our kinetic model in the light of $e$-$i$ collisions.}

\section{Simulation results\protect\label{sec:Simulation-results}}

\subsection{The premise}

Though technically it should be possible to model a self-consistent
periodic fluctuation of charge in external debris embedded in a
plasma, in reality it is much harder to model it mathematically as
it would naturally require calculation of collision cross-sections
to exactly determine the electron and ion currents to the debris,
which may not obey the much used orbit motion limited (OML) theory,
usually used to calculate the currents. It may be further complicated
by the extended and arbitrary shape of the debris. Besides, a self-consistent
fluctuation will necessarily require a coupling of a circuit-like differential
equation with the partial differential equations of the fluid model
for the plasma, so that we shall require a charging
equation like
\begin{equation}
\frac{dQ_{{\rm d}}}{dt}=I_{i}(t)+I_{e}(t)+I_{{\rm wave}}(t)+\cdots,\label{eq:charging}
\end{equation}
to be coupled to the plasma fluid equations, where $Q_{d}$ is the
charge on the debris and $I_{i,e,{\rm wave}}$ are the currents to
the debris due to plasma ions, electrons, and the IAW. For a fluid
simulation, the most challenging part would be to model the ``wave''
current. On the other hand, a PIC simulation may naturally model the
$I_{{\rm wave}}$ term through kinetic effects, but it does not natively
compute the above charging equation dynamically and would require
some kind of approximation with a sheath and potential model around
the debris. Besides, it would also require much higher grid resolution
for realistic calculation of these currents to the debris. As a result,
one would ideally need a hybrid approach of fluid and PIC methods
to model these fluctuations self-consistently.

In order to avoid these complexities, we have chosen to model the
periodic fluctuation of the charge on external debris through an ``external
forcing'' equation for the debris charge density $\rho_{{\rm ext}}$
\begin{equation}
 \rho_{{\rm ext}}(x,t)=\rho_{0}(x){\cal F}(t),
\end{equation}
where ${\cal F}(t)$ is the time-dependent part. This model can be
seamlessly incorporated into a fluid as well as a PIC model and the
results can be very efficiently compared. Besides, the controllable
external forcing-like equation allows us to have a finer control on
the fluctuation parameters and helps us isolate different nonlinear
phenomena through controlled tuning. In what follows, we shall present
results from two different simulation methods -- PIC simulation and
fluid simulation with a flux-corrected transport solver, which justifies
these observations. 

 {Our primary results are due to the PIC simulation as it incorporates
all the kinetic effects and full nonlinearities and also model a realistic plasma situation. 
The FCT model is a reduced fluid model, results from which should largely support
the PIC simulation results.}

\subsection{The plasma model}

The 1-D $e$-$i$ plasma system with Boltzmannian electrons is modeled
with the following equations with an exclusive time-dependent external
charge density for the debris,
\begin{eqnarray}
\frac{\partial n_{i}}{\partial t}+\frac{\partial}{\partial x}(n_{i}v_{i}) & = & 0,\label{eq:cont}\\
\frac{\partial v_{i}}{\partial t}+v_{i}\frac{\partial v_{i}}{\partial x} & = & -\frac{1}{m_{i}n_{i}}\frac{\partial p_{i}}{\partial x}-\frac{e}{m_{i}}\frac{\partial\phi}{\partial x},\\
n_{e} & = & n_{0}\exp\left(\frac{e\phi}{T_{e}}\right),\\
\epsilon_{0}\frac{\partial^{2}\phi}{\partial x^{2}} & = & e(n_{e}-n_{i})-\rho_{{\rm ext}}(t;x-v_{d}t),\label{eq:pois}
\end{eqnarray}
{where $n_i$ and $n_e$ are the ion and electron number densities respectively, $v_i$ is the ion fluid velocity, $p_i$ is the ion pressure, $\phi$ is the electrostatic potential, $m_i$ is the electron mass, $e$ is the electronic charge, $\epsilon_0$ is the vacuum permittivity and $T_e$ is the electron temperature. The last term in the
Poisson equation, $\rho_{ext}$ is the term due to external charged debris which
is moving with respect to the bulk plasma at a constant velocity $v_{d}$.
Note that depending on whether the debris is charged positively or negatively,
$\rho_{{\rm ext}}\lessgtr0$. The pressure for ions
is treated as polytropic
\begin{equation}
p_{i}\propto n_{i}^{\gamma},
\end{equation}
where $\gamma$ is the ratio of specific heats.} The normalized model can be written
as
\begin{eqnarray}
\frac{\partial n_{i}}{\partial t}+\frac{\partial}{\partial x}(n_{i}v_{i}) & = & 0,\label{eq:cont}\\
\frac{\partial v_{i}}{\partial t}+v_{i}\frac{\partial v_{i}}{\partial x}+\gamma\sigma n_{i}^{\gamma-2}\frac{\partial n_{i}}{\partial x} & = & -\frac{\partial\phi}{\partial x},\label{eq:mom}\\
n_{e} & = & e^{\phi},\label{eq:boltz}\\
\frac{\partial^{2}\phi}{\partial x^{2}} & = & n_{e}-n_{i}-\rho_{{\rm ext}}(t;x-v_{d}t),\label{eq:pois1}
\end{eqnarray}
where the densities are normalised by the equilibrium plasma density
$n_{0}$, length is normalized by electron Debye length $\lambda_{D}$,
time is normalized by the inverse of ion-plasma frequency $\omega_{pi}$,
plasma potential $\phi$ is normalized by $(T_{e}/e)$. The temperature
is expressed in energy units and velocity is normalized by the
ion-acoustic speed $c_{s}=\sqrt{T_{e}/m_{i}}$. The quantity $\sigma$
is the ratio of ion to electron temperature and is usually $\ll1$.
In the above equations, the external charge density $\rho_{{\rm ext}}$
is also normalized accordingly.

We consider the external charge density to be either positive definite
or negative definite, so that the charge in particular debris remains
same (either positive or negative) while oscillating in time.

\subsection{Kinetic model -- PIC simulation}

\subsubsection{The PIC simulation}
 {It is important for the general reader that we explain the basics of PIC simulation of plasma, though it is widely used and described in details in many good textbooks, especially one by Birdsall \citep{birdsall}. The dynamics modelled by PIC simulation is as per the Boltzmann-Vlasov (or simply Vlasov) equation \citep{chen-1974}
\begin{equation}
\frac{\partial f_{j}}{\partial t}+\bm{u}_{j}\cdot\nabla f_{j}+\frac{q_{j}}{m_{j}}\bm{E}\cdot\nabla_{u_{j}}f_{j}=0,\label{eq:vlasov}
\end{equation}
where $j=i,e$ stands for ions and electrons, $f_{j}$ are the respective velocity distributions, $\bm{u}_{j}$ are the respective particle velocities, in contrast to fluid velocity $v_i$ used in Eqs.(\ref{eq:cont}-\ref{eq:boltz}), and $\bm{E}$ is the plasma electric field. The operator $\nabla_{u_{j}}$ is the gradient operator in the respective velocity space. In presence of non-zero collisional momentum exchange, the right hand side of the above equation must be replaced by a collision term (which is zero in our case). The fluid equations given by Eqs.(\ref{eq:cont}-\ref{eq:boltz}) are nothing but differently weighted average of the Vlasov equation Eq.(\ref{eq:vlasov}). While the Vlasov equation is fully kinetic, the fluid model looses out the kinetic effects (those effects which involves individual particle properties) due to the averaging. Naturally, the fluid model is an approximation to the kinetic model.

The crux of the PIC model lies in its ability to model a large number of real-life plasma particles (of each kind) through a computational particle (also known as so-called \emph{super-particle\/}) \citep{birdsall}, which greatly reduces the computation time. Unless we consider a very strongly coupled plasma, the PIC model proves to be a very efficient method to model kinetic plasma phenomena \citep{birdsall}. The PIC model starts with distribution of the computational particles as per the given velocity distribution (which in our case, is a Maxwellian) on a grid, which is a 1D uniform grid in our case. Our particle distribution is spatially random and uniform. Once the particles are distributed, the electrostatic potential $\phi$ is calculated at the grid points through a nearest neighbour approximation \citep{birdsall,suniti}, which is then used to calculate the plasma electric field $\bm{E}$ at the grid points through the relation
\begin{equation}
\bm{E}=-\nabla\phi.\label{eq:ephi}
\end{equation}
This provides us with the force $\bm{F}$ at the particle positions through $\bm{F}=q\bm{E}$, which is again interpolated through a nearest neighbour scheme \citep{birdsall,suniti}. Once the force is obtained, the following equations are inverted subsequently to update new velocity $\bm{u}$ and then new position $\bm{x}$,
\begin{eqnarray}
\frac{d\bm{u}}{dt} &=& \frac{\bm{F}}{m},\\
\frac{d\bm{x}}{dt} &=& \bm{u}
\end{eqnarray}
and the cycle repeats.}

\subsection{Description of the simulation}

%%%%%%%%%%%%%%%%%%

In this section, we shall  {present the results of}  PIC simulation of the above-mentioned
$e$-$i$ plasma with a time-varying external charged debris. The
simulation is carried out with our well-tested \emph{hybird}-PIC-MCC
code \citep{suniti,suniti1,suniti3,mrid}, which has already been
successfully applied to several electrostatic problems including plasmas
with secondary electron emission, kinetic dust-ion-acoustic wave with
ion-Landau damping effect, and $e$-$i$ dusty plasma with external
charged debris.

The plasma is assumed to have an $e$-$i$ number density $n_{e,i}\sim10^{16}\,{\rm m}^{-3}$,
with electron temperature $T_{e}\sim1\,{\rm eV}$ and ion temperature
$T_{i}\sim0.01\,{\rm eV}$. The simulation box is of length $L\sim140\lambda_{D}\sim0.01\,{\rm m}$.  {The cell size for computation is $\sim6.25\times10^{-6}$ m which is equivalent to $\sim0.05\lambda_{D}$, enough to resolve fine-scale, Debye-scale fluctuations. The simulation is being performed with periodic boundary conditions. The simulation time and the box length is  adjusted, so that the nonlinearities in the neighbourhood of the debris are not affected by the boundaries.} The code can handle any arbitrary-shaped external charge distribution
of arbitrary strength which can be either positive or negative with
time-varying charge density. We can either make the debris move across
the plasma or vice versa -- the results being same in either case.
Computationally, as mentioned before, we model our time-varying charge
density with the following function
\begin{eqnarray} 
\rho_{{\rm ext}}(x,t) & = & \rho_{0}(x)[1+\tanh\{\Delta\sin(\nu t)\}]+\rho_{{\rm noise}}(t),\label{eq:profile}\\
\rho_0(x) & = & \tilde{\rho}_0e^{-x^2/w},
\end{eqnarray}
which is a {spatially} symmetric (Gaussian-like) charge distribution {$\rho_0(x)$, oscillating in time. 
The parameter  $\Delta$
is} the \emph{flatness} parameter of the time variation, $\nu$
{is} the oscillation frequency of the charge distribution, {and $w$ is the measure of spatial width of the distribution which is $\sim2\lambda_D$, with $\tilde{\rho}_0$ denoting the peak of the debris charge density. It should be noted that the value of $\tilde{\rho}_0$ \emph{must} be more than a critical value in order to sustain the debris-induced nonlinearity. In our case, with the present plasma parameters, the minimum $\tilde{\rho}_0^{\textrm{(min)}}\sim0.08$ with respect to the background charge density \citep{mrid}. If it is not enough, the debris will be overwhelmed by the inherent noise in the system and the debris-induced effects will diffuse within a few plasma periods.}
The larger
the value of $\Delta$ is, the steeper is the variation. Note that
the $\rho_{{\rm noise}}(t)$ part denotes a stochastic part of {$\rho_{{\rm ext}}(x,t)$},
which adds a degree of random-ness to the oscillation.

\begin{figure}[t]
\begin{centering}
\includegraphics[width=0.5\textwidth]{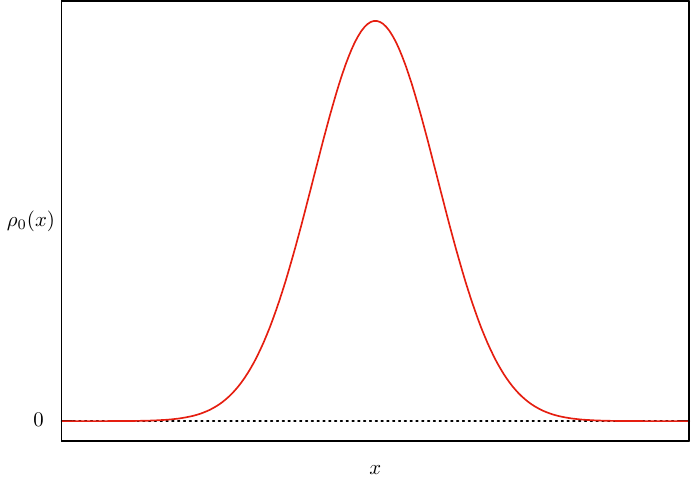}\hfill{}\includegraphics[width=0.49\textwidth]{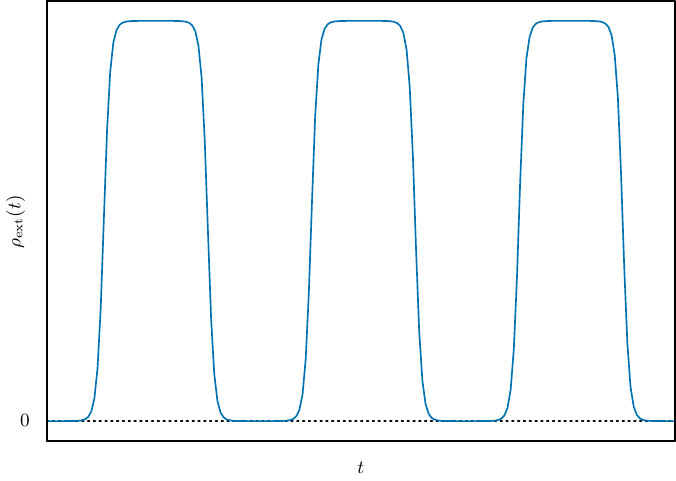}
\par\end{centering}
\caption{\protect\label{fig:modul}A schematic (not to scale) representation of the spatial distribution $\rho_0(x)$ (left) for the external charge  and temporal variation of the distribution (right). These representations are without stochastic noise.}
\end{figure}

\subsubsection{Charge modulation}
{In Fig.\ref{fig:modul}, we have schematically shown the spatial distribution of the external charge and its temporal variation (without noise). While the spatial distribution is a symmetric Gaussian, the distribution oscillates much like a top-hat function, oscillating between zero and some positive (or negative) value.

Or in other words, the charge distribution always remain positive or negative definite during the oscillation. The primary reason for choosing the oscillation is the way external debris get charged in the plasma. If we do not alter the debris properties (or plasma properties), the debris can get charged only in one way --  either positive or negative, not both. One can, however get external debris charged in both ways if the mechanism of charging becomes completely different over time like the lunar surface. We know that the Moon acts like an external entity in the flow of solar wind plasma. During day-time, the lunar surface gets charged positively due to the solar UV radiation inducing photoemission of electrons from the lunar surface, while during night-time, the surface gets negatively charged. But, in our case, this fluctuation of charge on the debris is due to inherent plasma oscillations where there is no way that the charging mechanism can get changed within the oscillation time.

If we go a bit deeper, we see that the whole process of charging of external entities in a plasma is due to the asymmetric response of plasma electrons and positive charges toward the external surface. So, we can say that the top-hat-like fluctuation is rooted in the asymmetric response of the plasma.}

%%%%%%%%%%%%%%%%%%%%%

\subsubsection{Modeling of the noise}

{We treat the noise as one which can naturally arise in a plasma due to
random fluctuations. If we denote the driving signal as $S(t)$, which
in our case may be a periodic signal due to plasma wave, the total
signal $T(t)$ can be thought to be composed of the signal $S(t)$
and a stochastic noise $R(t)$. However, we assume that the stochastic
fluctuation is purely \emph{environmental} i.e.\ it depends on the
signal $S(t)$, so that the total signal can be written as
\begin{equation}
T(t)=S(t)+R(t),
\end{equation}
where
\begin{equation}
R(t)=\begin{cases}
R_{0}(t), & \textrm{if}\,S(t)=0,\\
|\delta R(t)|S(t), & \textrm{if}\,S(t)>0,
\end{cases}\quad\delta R(t)<0.
\end{equation}
This model can be justified by considering the fact that even in the
absence of any driving signal, the plasma continues to have background
thermal fluctuations such as spontaneous Langmuir oscillations
\citep{yoon}. This explains the relation
\begin{equation}
T(t)=R_{0}(t),\quad\textrm{if}\,S(t)=0.
\end{equation}
On the other hand, when there is a driving signal (such as in our
case), the plasma responds nonlinearly, which can be through nonlinear
wave-particle interactions (viz.\ nonlinear Landau damping) \citep{nastac,mallet}.
As this response is supposed to be dependent on the driving signal,
we have
\begin{equation}
T(t)=S(t)[1-|\delta R(t)|].
\end{equation}
Regarding the strength of the stochastic component, we have assumed
it to be $R_{0}(t)\leq S(t)$ and $|\delta R(t)|\leq1$. Though there
is not any hard justification to this assumption, it is quite intuitive
that stochastic fluctuations are always supposed to be lower than
the driving signal, when the driving signal goes to zero and it always
attenuates the signal, whenever it is non-zero \citep{nastac,mallet}.}

%%%%%%%%%%%%%%%%%%%%%

\subsubsection{Chaotic dynamics}
{We are now going to discuss the chaotic dynamics induced by  different oscillating external charge distributions.}

The variable
data $(v_{i},n_{i})$ are extracted across the simulation domain as
independent time series, which are then examined with Wolf's algorithm
for possible chaotic nature. It is now well known that the Wolf's algorithm
\citep{wolf} applies Taken's theorem \citep{takens} to construct
an $m$-dimensional phase space from a scalar time series $\{x_{t}\}$,
\begin{equation}
\bm{X}_{t}=\left[x_{t},x_{t+\tau},x_{t+2\tau},\cdots,x_{t+(m_{{\rm PIC}}-1)\tau}\right],
\end{equation}
with $m_{{\rm PIC}}$ as the embedding dimension and $\tau$ as the
time delay. The algorithm continuously measures the divergence between
two nearest neighbour (forward in time)
\begin{equation}
\delta(t)=\left\Vert \bm{X}_{t}-\bm{X}'_{t}\right\Vert ,
\end{equation}
where $\bm{X}'_{t}$ is $\bm{X}_{t}$'s nearest neighbour (in time).
The so-called Lyapunov exponent \citep{lyapbook}, which is nothing
but the exponential divergence of the nearest trajectories, is then
calculated over \emph{valid} divergence segments as
\begin{equation}
\lambda_{j}=\frac{1}{\Delta t}\,\ln\left[\frac{\delta(t+\Delta t)}{\delta(t)}\right],
\end{equation}
where $\Delta t$ is the time difference taken to compute the divergence.
The maximal Lyapunov exponent is then calculated by
\begin{equation}
\lambda=\frac{1}{N}\sum_{j=1}^{N}\lambda_{j},
\end{equation}
where $N$ is the number of valid divergence segments. Any positive
$\lambda$ indicates chaos and the further $\lambda$ is away from
$0$, the stronger is the chaos. A typical rule of thumb is to have
$\lambda\gtrsim0.01$ for a definitive proof of chaos. In our case,
we have used {the variable $n_i$ for the scalar time series $\{x_i\}$ and} an embedding dimension $m_{{\rm PIC}}=4$ for our chaos
test.
\begin{figure}[t]
\begin{centering}
\includegraphics[width=0.5\textwidth]{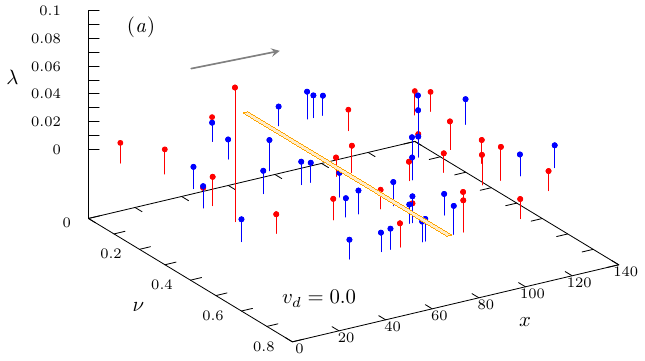}\hfill{}\includegraphics[width=0.5\textwidth]{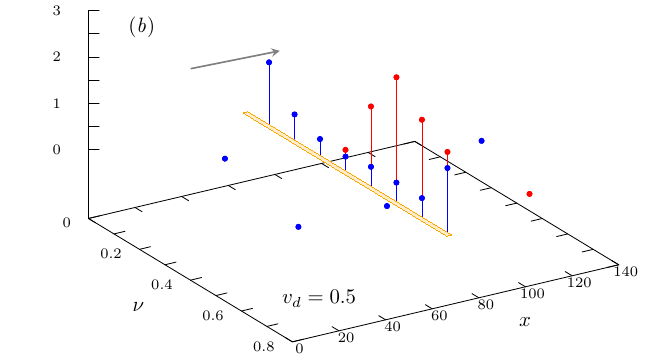}\\
\includegraphics[width=0.5\textwidth]{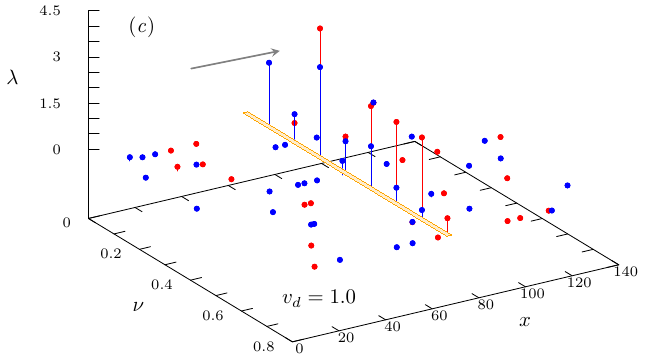}\hfill{}\includegraphics[width=0.5\textwidth]{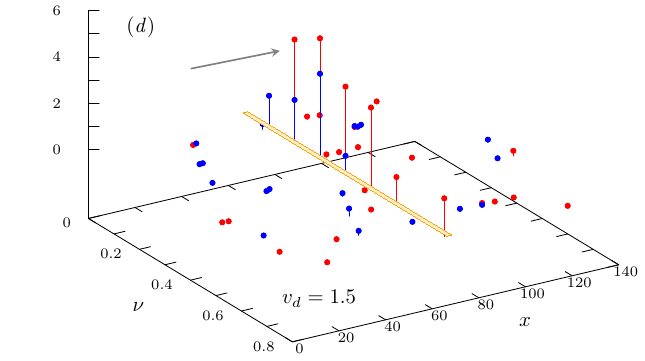}
\par\end{centering}
\caption{\protect\label{fig:Maximal-Lyapunov-exponent}Maximal Lyapunov exponent
$\lambda$, shown in the $x$-$\nu$ plane for a positively charged
external debris (shown as the orange strip). The blue and red points
indicate values of $\lambda$ with and without stochastic fluctuation.
The results are at time $\sim20\tau_{pi}$. The arrow indicates the direction of debris velocity.}
\end{figure}

\begin{figure}[t]
\begin{centering}
\includegraphics[width=0.5\textwidth]{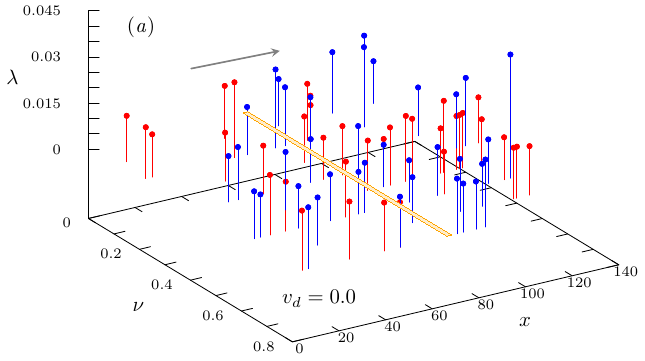}\hfill{}\includegraphics[width=0.5\textwidth]{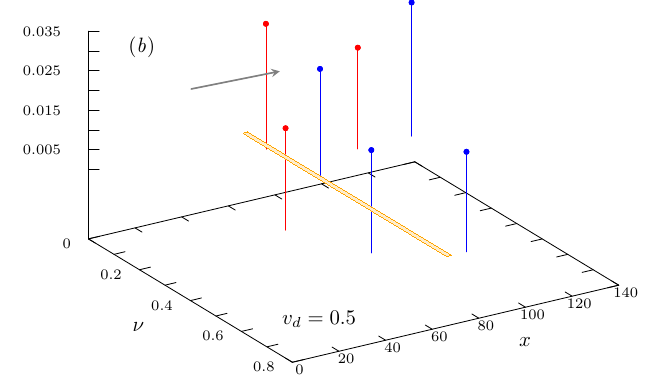}\\
\includegraphics[width=0.5\textwidth]{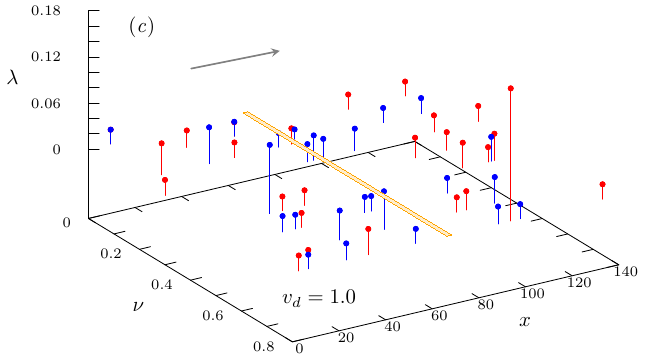}\hfill{}\includegraphics[width=0.5\textwidth]{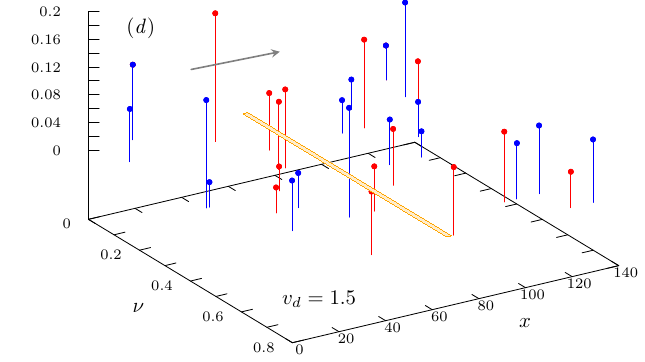}
\par\end{centering}
\caption{\protect\label{fig:Maximal-Lyapunov-exponent-1}Maximal Lyapunov exponent
$\lambda$, shown in the $x$-$\nu$ plane for a negatively charged
external debris (shown as the orange strip). The blue and red points
indicate values of $\lambda$ with and without stochastic fluctuation
as before. The results are at time $\sim20\tau_{pi}$.  The arrow indicates the direction of debris velocity.}
\end{figure}

The result of the simulation is shown in Figs.\ref{fig:Maximal-Lyapunov-exponent}
and \ref{fig:Maximal-Lyapunov-exponent-1}, where we have shown the
scatter plots of $\lambda$ in a $x$-$\nu$ plane. The position of
the debris is indicated by an orange-colored strip in the panels.
The position is expressed in terms of electron Debye length $\lambda_{D}$
and the frequency $\nu$ is expressed in terms of ion plasma frequency
$\omega_{pi}$. In order to help the perspective visualization of
the value of $\lambda$, we have added ``vertical drop-lines'' from
every point to the $z=0$ baseline so that the length of the `drop-line'
indicates the value of $\lambda$. All simulation results are at time
$\sim20\tau_{pi}$, where $\tau_{pi}$ is the plasma period $=\omega_{pi}^{-1}$.
The results for positively charged debris is shown in Fig.\ref{fig:Maximal-Lyapunov-exponent}
and that for negatively charged debris are shown in Fig.\ref{fig:Maximal-Lyapunov-exponent-1}.
As we can see from Fig.\ref{fig:Maximal-Lyapunov-exponent} that for
positively charged debris, increase in debris velocity definitely
induces quite stronger chaos in the oscillation, specifically at the
site of the debris. The average strength of the $\lambda$ in panel
(\emph{a}) is almost in the boundary line of chaos $\sim0.01$, which
is without any relative motion between the plasma and external debris
$(v_{d}=0)$. This should be contrasted with those in panels (\emph{b}),
(\emph{c}), and (\emph{d}), where we can see a definitive built up
of chaotic oscillations. It should also be noted that the site of
chaotic oscillation is mostly confined to the debris site. Besides,
the stochasticity of fluctuation is not seen to have any systematic
and correlated effect on $\lambda$, a situation which we shall explain
later in Section \ref{subsec:Effect-of-stochastic}.

In contrast to the positively charged debris, we can see a completely
different scenario in case of negatively charged debris from Fig.\ref{fig:Maximal-Lyapunov-exponent-1}.
In this case, we can see that the strength of the chaos is quite weak
as compared to the chaos induced by fluctuating positively charged
debris. Besides, the chaotic points are almost evenly distributed
across the $x$-$\nu$ plane as opposed to the earlier case where
the chaos is mostly confined to the debris site. As far as the effect
of stochastic fluctuation is concerned, it is similar to that of positively
charged debris.

\begin{figure}[t]
\begin{centering}
~~\includegraphics[width=0.51\textwidth]{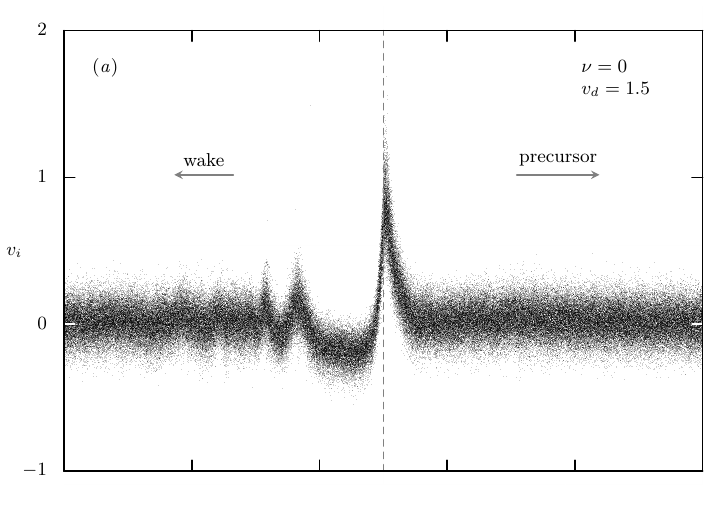}~\hskip-13pt~\includegraphics[width=0.51\textwidth]{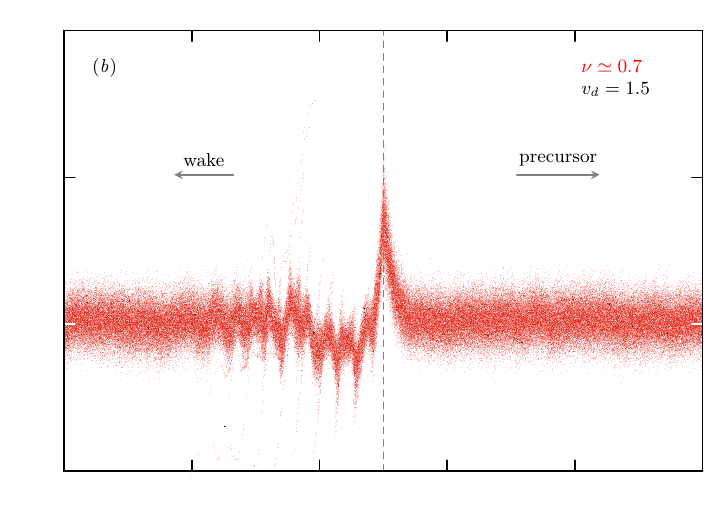}\\
~\vskip-36pt~\\
\includegraphics[width=0.518\textwidth]{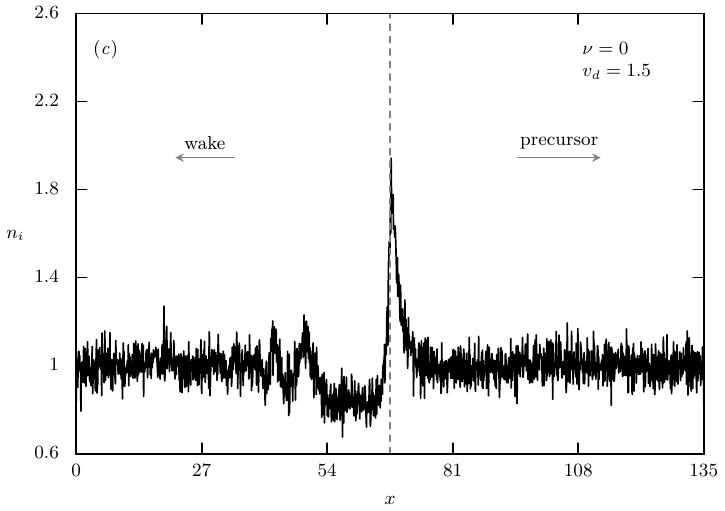}~\hskip-16pt~\includegraphics[width=0.518\textwidth]{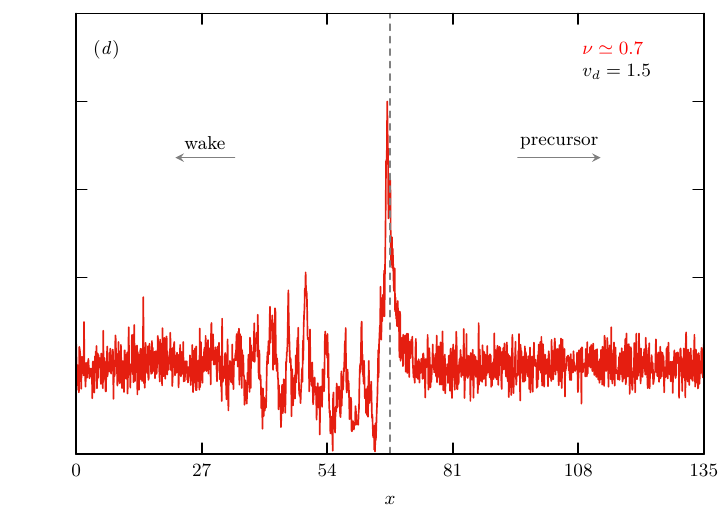}
\par\end{centering}
\caption{\protect\label{fig:Ion-phase-space}Ion phase space plots and ion
density plots are shown in panels (\emph{a}), (\emph{b}) and (\emph{c}), (\emph{d}) for positively charged external debris. The
left hand plots (black colored) are without any fluctuation of the
debris charge and the right hand plots (red colored) are for periodic
fluctuation of the debris charge. The value of the fluctuation frequency
$\nu$ is given in each panel.}
\end{figure}

\begin{figure}[t]
\begin{centering}
\includegraphics[width=0.5\textwidth]{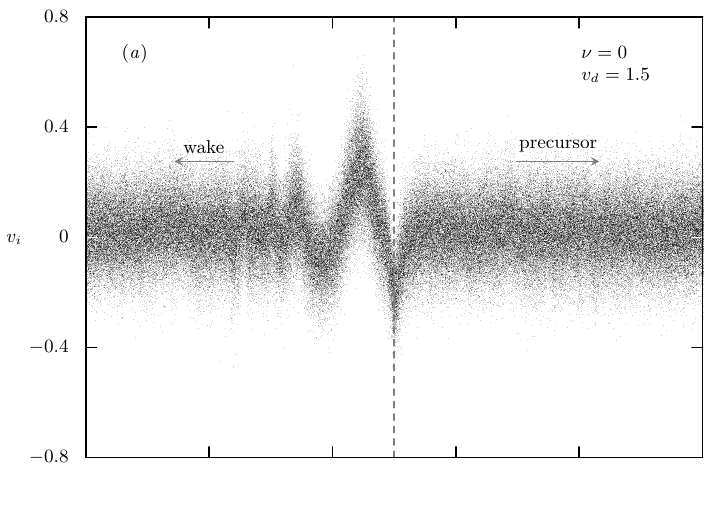}\hfill{}\includegraphics[width=0.5\textwidth]{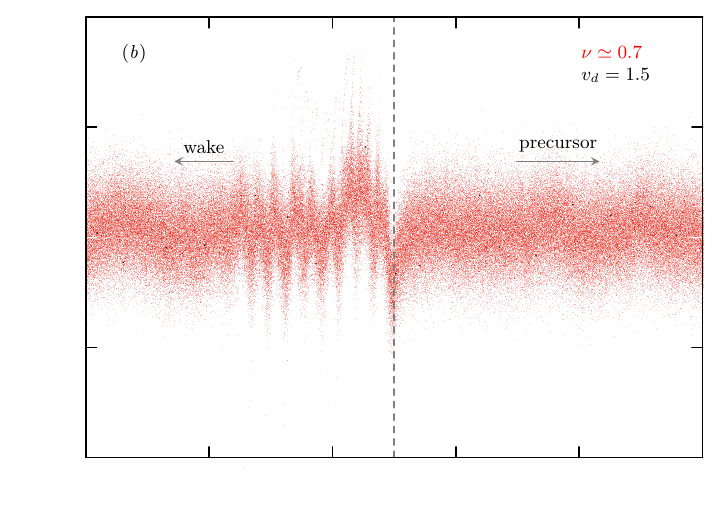}\\
~\vskip-36pt~\\
~\hskip-0pt\includegraphics[width=0.495\textwidth]{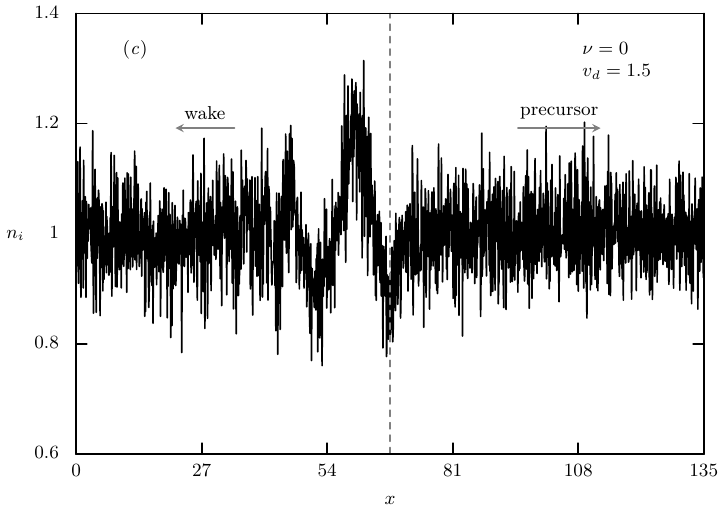}~\hskip-5pt~\includegraphics[width=0.495\textwidth]{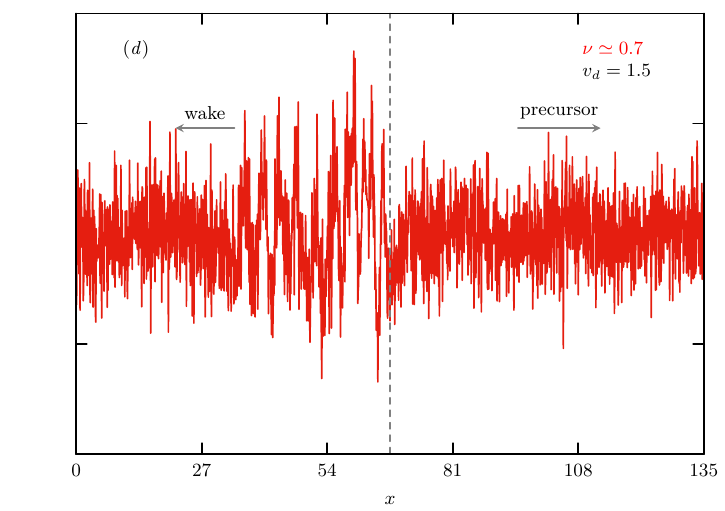}
\par\end{centering}
\caption{\protect\label{fig:Ion-phase-space-1}Ion phase space plots and ion
density plots for negatively charged external debris. All other particulars
are same as in Fig.\ref{fig:Ion-phase-space}. The value of the fluctuation
frequency $\nu$ is given in each panel.}
\end{figure}

The corresponding ion phase space and ion density plots are shown
in Fig.\ref{fig:Ion-phase-space} for positive debris and in Fig.\ref{fig:Ion-phase-space-1}
for negative debris. All frames of Figs.\ref{fig:Ion-phase-space}
and \ref{fig:Ion-phase-space-1} are drawn in the rest frame of the
debris, which lies in the middle of the simulation window denoted
by a dashed vertical line in each panel. The ion velocity is normalized
to the ion-acoustic speed and the ion density is normalized by its
equilibrium value. The length is normalized by electron Debye length
$\lambda_{D}$. All these plots are without any stochastic fluctuation
component. What we can see from the plots is that, the periodic charge
fluctuation of the external debris has a minimal effect when the debris
charge is negative, also manifested by weak chaos, as shown before. 


\subsection{Fluid model -- FCT simulation }

In this section, we shall try to model the fluctuation-induced chaos
from a fluid perspective with flux-corrected transport simulation
\citep{sarkar2023response,sarkar2025exploring}. Here, we use a multi-fluid
FCT (\emph{m}FCT) code based on Boris's original algorithm \citep{boris1973}
with Zalesak's flux limiter \citep{zalesak}, which we use to solve
Eqs.(\ref{eq:cont}-\ref{eq:pois}) with time-varying $\rho_{{\rm ext}}$.
This code has also been extensively used to study debris-induced plasma
wave phenomena \citep{sarkar2025exploring}.

The FCT formalism requires the hydrodynamic equations Eqs.(\ref{eq:cont}-\ref{eq:mom})
to be put in the form of a generalized continuity equation,
\begin{equation}
\frac{\partial f}{\partial t}=-\frac{\partial}{\partial x}(fv_i)+\frac{\partial s}{\partial x},\label{eq:fct}
\end{equation}
where $f=(n_{i},v_{i})$ is the physical quantity to be solved, {$(fv_i)$}
is the corresponding flux, and $s$ is the source term. The external
charged debris is entered through the Poisson equation Eq.(\ref{eq:pois})
as before which is to be solved at each time step along with Eq.(\ref{eq:fct}).
The external charged debris is modeled with
the same profile as used in the PIC simulation, given by Eq.(\ref{eq:profile}).

We now present a full-spectrum Lyapunov exponent (LE) $\lambda_{k}$
analysis of the velocity profile on the debris site, as obtained from
FCT simulation. The algorithm to calculate Lyapunov exponent is based
on Wolf's algorithm \citep{wolf} for an embedding dimension $m_{{\rm FCT}}=6$.
\begin{equation}
\lambda_{1}\ge\lambda_{2}\ge\cdots\ge\lambda_{m_{\rm FCT}},\quad m_{{\rm FCT}}=6,
\end{equation}
for a scalar time series $v_{i}$, at the site of the debris, resulted
out of the fluid simulation. We start by constructing the delay vectors
\begin{equation}
\vec{f}_{j}=\left[f(t_{j}),f(t_{j}+\delta t),\dots,f(t_{j}+(m_{{\rm FCT}}-1)\delta t)\right]\,{\rm in}\,\mathbb{R}^{6},
\end{equation}
required for phase-space reconstruction. The embedding dimension $m_{{\rm FCT}}=6$
is further verified through Kaplan-Yorke (KY) dimensional analysis
\citep{kaplan} of the embedding dimension, 
\begin{equation}
D_{{\rm KY}}=j+\frac{\sum_{k=1}^{j}\lambda_{k}}{|\lambda_{k+1}|},\quad\sum_{k=1}^{j}\lambda_{k}>0,
\end{equation}
where the summation is over all non-zero $\lambda_{k}$. This yields
KY fractal dimension $D_{{\rm KY}}\approx5$ for our $\lambda_{k}$s
which conforms to the general guideline $m_{{\rm FCT}}>D_{{\rm KY}}$,
an over-conservation as compared to Taken's theorem \citep{takens}
that in general requires $m_{{\rm FCT}}>2D_{{\rm KY}}$. While calculating
the Lyapunov spectrum, we have used a Theiler window \citep{lyapbook}
to avoid temporal correlation and tracked the divergence of the nearby
trajectories using a linearized approximation. A Gram-Schmidt orthonormalization
(i.e.\ QR decomposition) \citep{qr} is applied in each time step
to the evolving tangent vectors and the LEs are calculated finally
by
\begin{equation}
\lambda_{k}=\frac{1}{T}\sum_{k=1}^{N}\log R_{kk}^{(p)}.
\end{equation}
In the above expression, $R$ is the upper-triangular matrix as obtained
from the QR decomposition, $p$ is the index of time step at which
the decomposition is carried out, $N$ is total number of such steps,
and $T$ is the physical time $T=\sum_{k=1}^{N}\Delta t_{k}$ with
$\Delta t$ being the time interval (of the time series) between each
decomposition step.

Note that the reason for the embedding dimension for FCT simulation
$(m_{{\rm FCT}}=6)$ being higher than that for PIC simulation $(m_{{\rm PIC}}=4)$
can be understood by considering the presence of ``noise'' in PIC
simulation, which is almost absent in FCT simulation as it numerically
solves the model equations exactly. On the other hand, the ``noise''
in PIC simulation scales as inverse of the square root of total number
of computational particles \citep{picnoise}
\begin{equation}
\textrm{PIC Noise}\propto\frac{1}{\sqrt{N_{p}}},
\end{equation}
where $N_{p}$ is the total number of computational particles used
in the PIC simulation. As computational cost per PIC cycle is $\propto N_{p}$,
a decrease in ``noise'' by a factor of $2$ increases computational
cost by a factor of $4$, making the option of increasing $N_{p}$
not a viable one. Naturally, the FCT simulation can detect more fine-scale
fluctuations whereas PIC simulation picks up more coarse-grained fluctuations
which results in a low-dimensional dynamical system for PIC simulation.

\begin{figure}[t]
\begin{centering}
\includegraphics[width=0.5\textwidth]{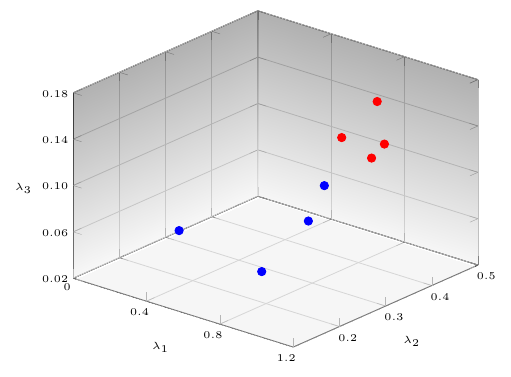}
\par\end{centering}
\caption{\protect\label{fig:Full-spectrum-LEs}Full spectrum LEs calculated
from the time series obtained from the FCT simulation.}
\end{figure}

The results of this analysis is shown in Fig.\ref{fig:Full-spectrum-LEs},
which shows the full spectrum LE. In the figure, the axes are the
three dominant LEs $(\lambda_{1,2,3})$ and the red and blue circles
respectively denote the cases $\rho_{{\rm ext}}>0$ and $<0$. Apart
from the fact that all $\lambda_{1,2,3}>0$ which signifies chaos,
the accumulation of the red circles at the upper portion of the figure
indicate a considerably stronger chaos for $\rho_{{\rm ext}}>0$ than
$<0$. The four cases corresponding to four circles in the figure
(red and blue) are for different debris velocities $v_{d}=0.2,0.5,1.0$
and $1.5$ (the numerical code used to calculate the full-spectrum
LE is available).

\begin{figure}[t]
\begin{centering}
\includegraphics[width=0.5\textwidth]{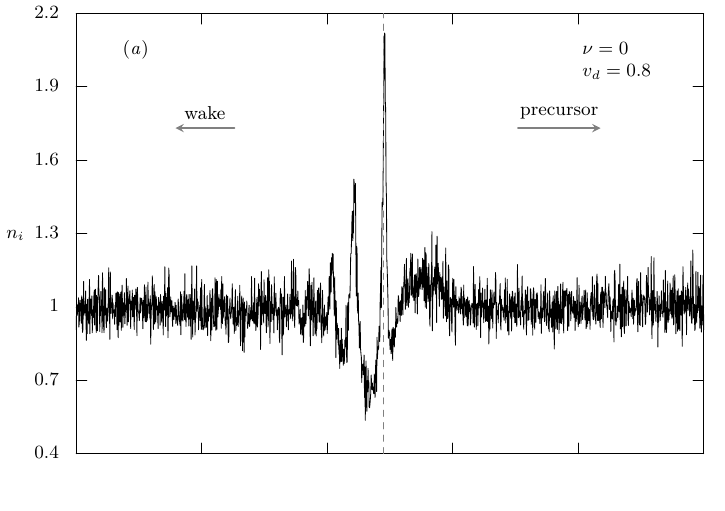}\hfill{}\includegraphics[width=0.5\textwidth]{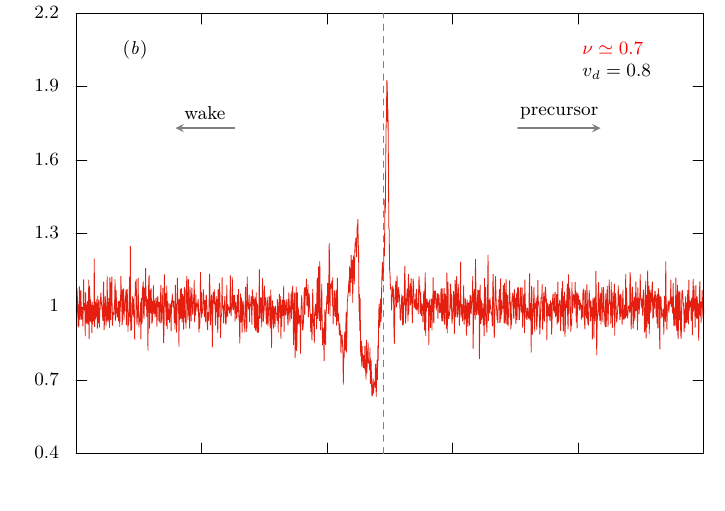}\\
~\vskip-36pt~\\
~\hskip5pt\includegraphics[width=0.48\textwidth]{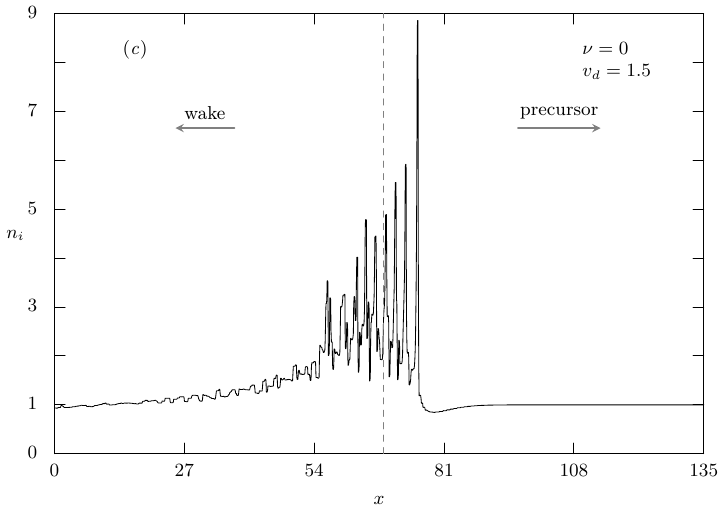}~\hskip5pt\includegraphics[width=0.48\textwidth]{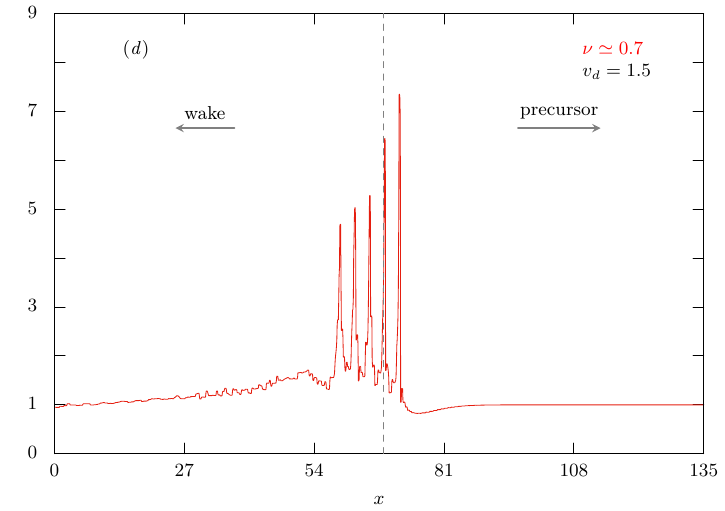}
\par\end{centering}
\caption{\protect\label{fig:Suppression-of-density}Suppression of density
oscillations with periodic charge fluctuation of the negatively charged
debris. The top panels (\emph{a}), (\emph{b}) are obtained from PIC
simulation while the bottom panels (\emph{c}), (\emph{d}) are obtained
from FCT simulation. The panels on the left are without any charge
fluctuation and those on the right are with charge fluctuation. {The vertical dashed line denotes the position of the debris.}}
\end{figure}

\subsection{Nonlinear Landau damping?}

We now present a combined result of PIC and FCT simulations which
point out a phenomenon similar to what is known as ``nonlinear Landau
damping'' \citep{nlandau}. In both cases, we observe a suppression
of oscillations when $v_{d}\sim c_{i}$, where $c_{i}=\sqrt{T_{e}/m_{i}}$
is the phase velocity of the ion-acoustic oscillations. However, the
fundamental physics behind this observation is slightly different
in both cases. 
\begin{figure}[t]
\begin{centering}
\includegraphics[width=0.52\textwidth]{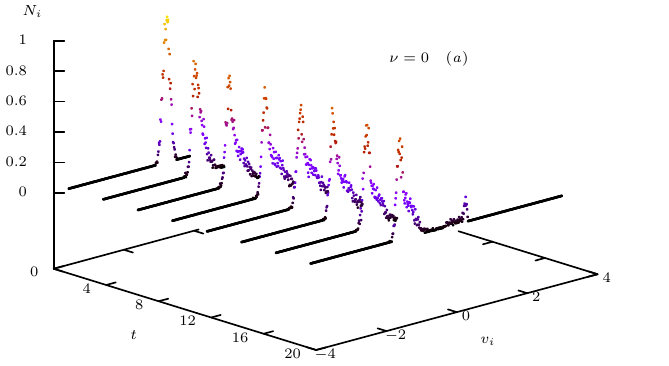}~\hskip-24pt~\includegraphics[width=0.52\textwidth]{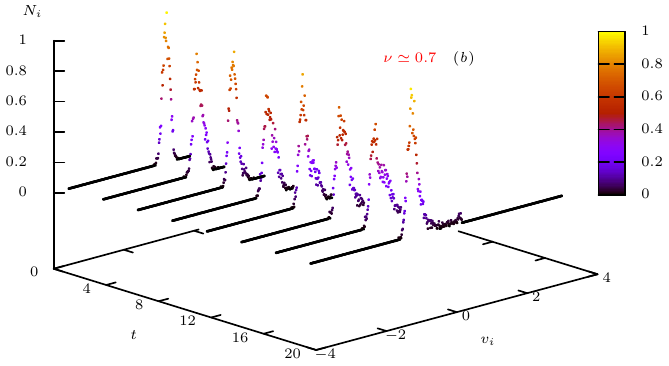}\\
~\vskip-12pt~\\
\includegraphics[width=0.5\textwidth]{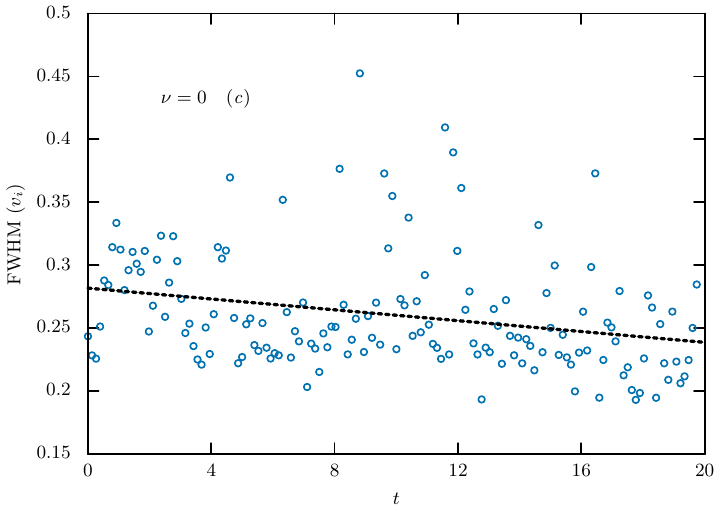}\includegraphics[width=0.5\textwidth]{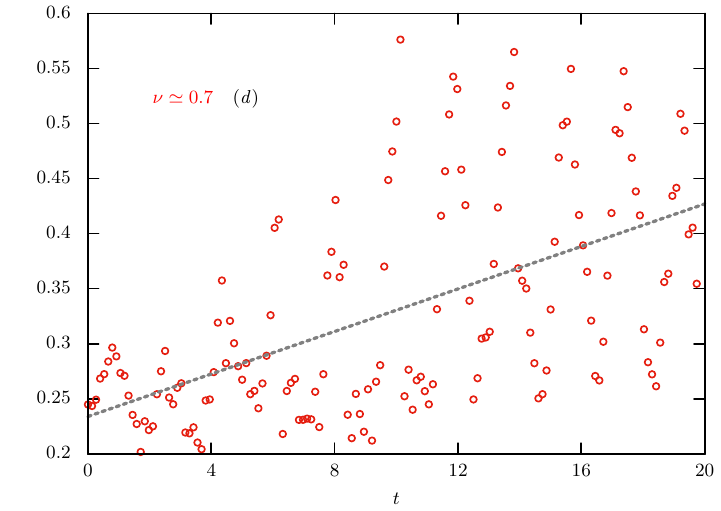}
\par\end{centering}
\caption{\protect\label{fig:Flattening-of-the}Flattening of the ion VDF due
to {charge fluctuation of the negatively charged debris.} While the top panels (\emph{a}) and (\emph{b}) display
respectively the VDFs calculated from the PIC simulation without $(\nu=0)$
and with $(\nu\sim0.7)$ fluctuation, the bottom panels (\emph{c}) and
(\emph{d}) show the numerically calculated FWHMs for the corresponding
VDFs. The panel (\emph{d}) on the right clearly shows that the VDF
is flattening due to fluctuation.}
\end{figure}

Consider an external charged debris which is moving with a velocity
$v_{d}$ with a potential profile $\phi(x-v_{d}t)$, modulated by
a time-varying charge $Q(t)$. The plasma responds to this debris
field through ion-acoustic oscillations with a phase velocity $c_{i}$.
So, when $v_{d}\sim c_{i}$, we have a resonant energy transfer between
the background plasma and the charged debris. However, the oscillating
charge of the debris is going to introduce another frequency scale
$\sim\nu$. As a result, the Doppler-shifted frequency of the oscillation
(as seen by the debris) is given by
\begin{equation}
\omega_{{\rm forcing}}=\nu+kv_{d}.
\end{equation}
Thus, we have a resonance condition, when $\omega_{{\rm forcing}}\simeq\omega_{{\rm IA}}=kc_{i}$,
or
\begin{equation}
v_{d}\simeq c_{i}-\frac{\nu}{k}.\label{eq:res0}
\end{equation}
We can expect a strong wake when a resonance occurs and destructive
interference when it becomes out of phase,
\begin{equation}
|\nu+k(v_{d}-c_{i})|\gtrsim0,\label{eq:res}
\end{equation}
reaching a maximum of $\pi$ when they become completely out of phase
by a turnover time of $\tau_{\pi}$
\begin{equation}
\tau_{\pi}|\nu+k(v_{d}-c_{i})|\sim\pi.\label{eq:res1}
\end{equation}
It is then more likely that `anti-resonance' occurs via relations
(\ref{eq:res}) and (\ref{eq:res1}) rather than a resonance condition
occurs, which has to be exactly via relation (\ref{eq:res0}). So,
we expect a suppression of oscillations to occur, on average, due
to the periodic variation of debris charge. This is the reason behind
the observation of suppression of oscillation seen in the FCT simulation
results. However, as the PIC simulation is expected to also incorporate
the effect of NLLD, we perform an additional check on the ion velocity
distribution function (VDF) as obtained from PIC simulation. Though
there are no definitive proofs of NLLD in general, for collision-less
electrostatic plasma such as ours, a flattening of the velocity distribution
can be attributed to NLLD. 

This effect is shown in Fig.\ref{fig:Suppression-of-density} for
negatively charged debris, both as obtained from PIC and FCT simulations.
However, this resonant suppression of oscillation occurs at $v_{d}\lesssim c_{i}$
for PIC simulation and at $v_{d}\gtrsim c_{i}$ for FCT simulation
as can be seen from the figure. This discrepancy can be explained
from the fact that the PIC simulation captures the full kinetic response
including wave-particle resonance and phase space trapping while the
FCT simulation captures only the macroscopic fluid-like wave-wave
response and lacks phase mixing. In Fig.\ref{fig:Flattening-of-the},
we have shown the ion VDF obtained from PIC simulation data with and
without debris charge fluctuation, which clearly show the flattening
of the VDF with fluctuation. The suppression of oscillations combined
with the definitive proof of flattening of the ion VDF in PIC simulation
strongly supports the fact that oscillations undergo NLLD due to charge
fluctuation in the case of negatively charged debris. We note here
that this flattening of ion VDF is not seen in case of fluctuating
positively charged debris. {In the above analysis, we have used $v_d=0.8$.}

\section{Theoretical analysis\protect\label{sec:Theoretical-analysis}}

\subsection{Torus breakdown of a chaotic plasma model}

In this section, we try to understand the theory behind the chaotic
oscillation observed in the simulations and try to understand some
of the observations related to stochastic fluctuations and why the
properties of chaos for positively and negatively charged debris are
fundamentally different. In particular, we shall discuss how the fluctuation
of charge density of a debris may force the system to undergo torus
breakdown \citep{greene,torus1,torus2} and become chaotic.

In order to facilitate a dynamical analysis, we make a coordinate
transformation to the moving frame of the debris with the introduction
of a scaled time $\tau=t-x/v_{d}$, so that in terms of the scaled
variable we have $\partial/\partial t\to\partial/\partial\tau$ and
$\partial/\partial x\to-v_{d}^{-1}\partial/\partial\tau$ \citep{suniti3}.
With this transformation, our plasma model represented by Eqs.(\ref{eq:cont}-\ref{eq:pois1})
becomes
\begin{eqnarray}
\frac{\partial n_{i}}{\partial\tau}-\frac{1}{v_{d}}\frac{\partial}{\partial\tau}(n_{i}v_{i}) & = & 0,\label{eq:cont-1}\\
\frac{\partial v_{i}}{\partial\tau}-\frac{v_{i}}{v_{d}}\frac{\partial v_{i}}{\partial\tau}-\frac{\gamma\sigma}{v_{d}}n_{i}^{\gamma-2}\frac{\partial n_{i}}{\partial\tau} & = & \frac{1}{v_{d}}\frac{\partial\phi}{\partial\tau},\label{eq:phi}\\
\frac{1}{v_{d}^{2}}\frac{\partial^{2}\phi}{\partial\tau^{2}} & = & n_{e}-n_{i}-\rho_{{\rm ext}}(\tau).
\end{eqnarray}
Integration of Eq.(\ref{eq:cont-1}) gives us
\begin{equation}
n_{i}=\frac{v_{d}-v_{0}}{v_{d}-v_{i}},\label{eq:ni}
\end{equation}
where $v_{0}$ is the velocity of the ions at infinity (or far away
from the debris). In the above integration, we have also used the
boundary condition that at $\infty$, $n_{i}\to1$. Note that for
the above formalism to work, we must have $v_{d}\lessgtr(v_{0},v_{i})$
or else the density would be negative, which is unphysical. Without
loss of any generality we can assume $\gamma=2$ which helps us to
integrate Eq.(\ref{eq:phi}) to have
\begin{equation}
\phi=\frac{1}{2}\left(v_{0}^{2}-v_{i}^{2}\right)-v_{d}(v_{0}-v_{i})+2\sigma\left(1-\frac{v_{d}-v_{0}}{v_{d}-v_{i}}\right),
\end{equation}
where we have used the expression given by Eq.(\ref{eq:ni}). Inserting
the above expressions in Poisson equation, we have the following nonlinear
second order differential equation in $v_{i}$,
\begin{equation}
\ddot{v}_{i}A+\dot{v}_{i}^{2}B+C=0,\label{eq:combined}
\end{equation}
where the `$\,\dot{}\,$' refers to derivative in $\tau$ and
\begin{eqnarray}
A & = & \frac{1}{v_{d}^{2}}\left(\mu_{i}-2\sigma\frac{\mu_{0}}{\mu_{i}^{2}}\right),\\
B & = & -\frac{1}{v_{d}^{2}}\left(1+4\sigma\frac{\mu_{0}}{\mu_{i}^{3}}\right),\\
C & = & \rho_{{\rm ext}}(\tau)+\frac{\mu_{0}}{\mu_{i}}-e^{\phi}
\end{eqnarray}
where $\mu_{i}=v_{d}-v_{i}$ and $\mu_{0}=v_{d}-v_{0}$. Note that
Eq.(\ref{eq:combined}) is a highly nonlinear non-autonomous equation.

\begin{figure}[t]
\begin{centering}
\includegraphics[width=1\textwidth]{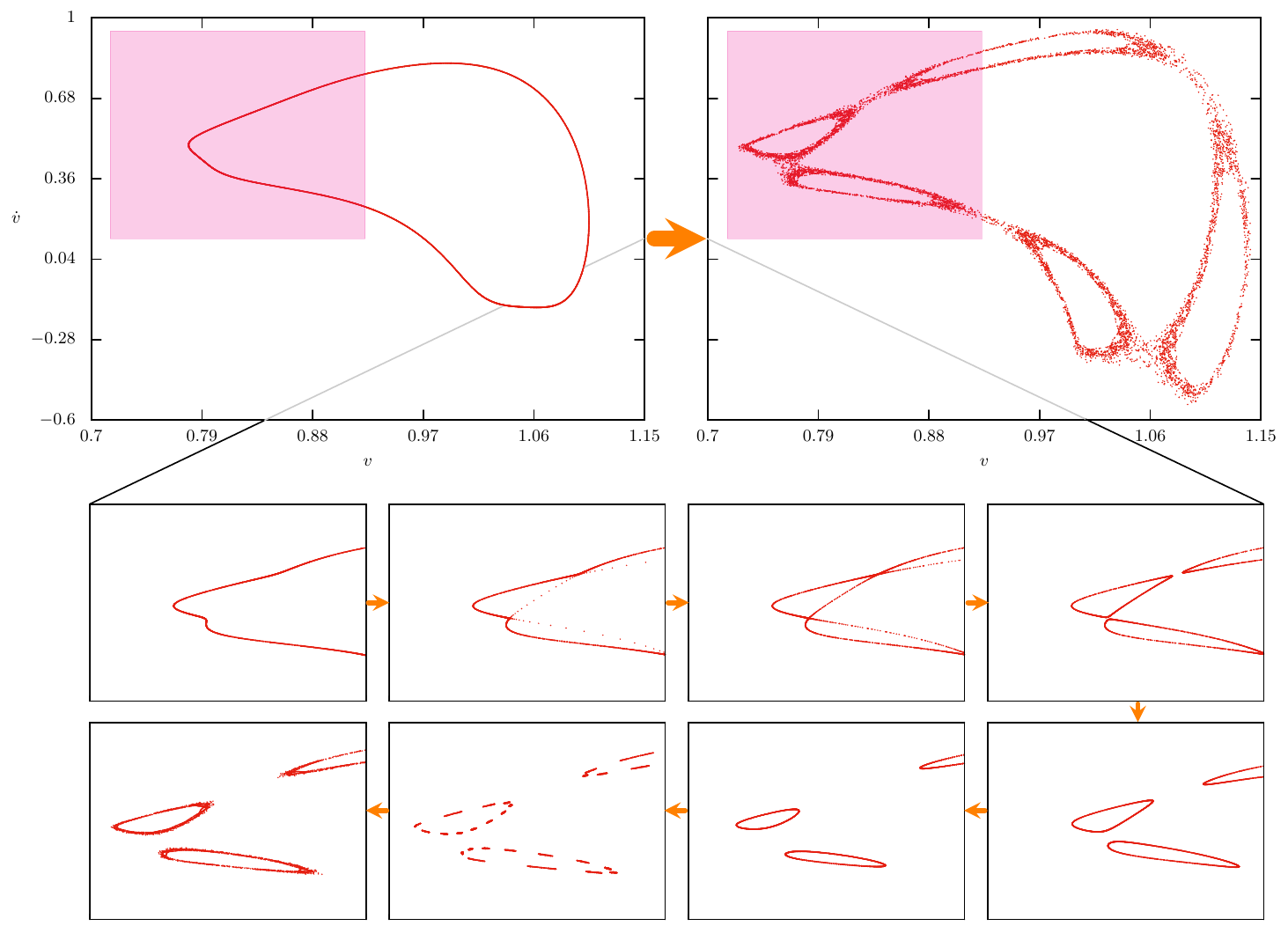}
\par\end{centering}
\caption{\protect\label{fig:Poincar-plots.}Poincar\'e plots of one transition
phase from periodic to a chaotic state through torus breakdown. The
initial (periodic) state is shown at the left top and the final chaotic
state is shown at the right top panels. The arrow and the blown up
transition phases are shown in the lower smaller panels.}
\end{figure}

\begin{figure}[t]
\begin{centering}
\includegraphics[width=0.5\textwidth]{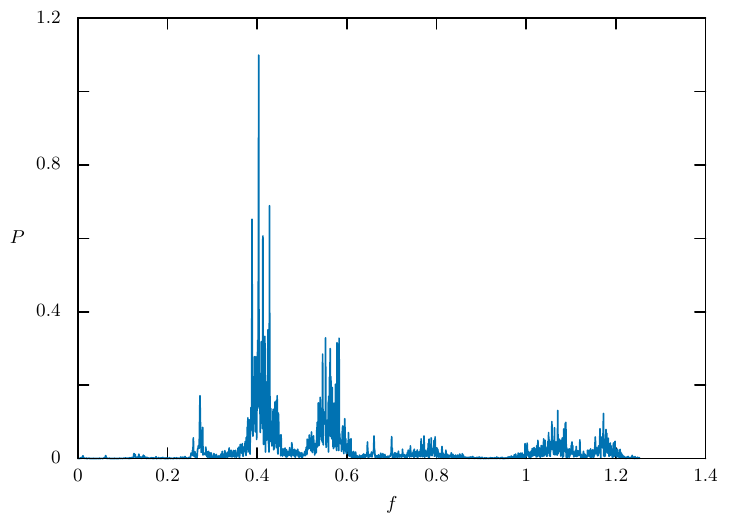}\hfill{}\includegraphics[width=0.5\textwidth]{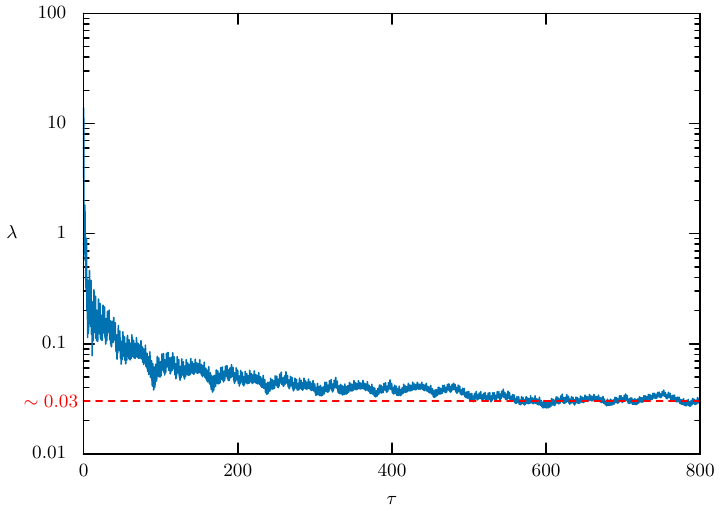}
\par\end{centering}
\caption{\protect\label{fig:Power-spectrum-density}Power spectrum density
at transition and maximal Lyapunov exponent at the final chaotic state.}
\end{figure}

\subsubsection{Torus breakdown}

As we have mentioned that the system is highly nonlinear, it can undergo
a chaotic transition with the variation of different control parameters.
The fluctuation frequency $\nu$ of the charge density of the external
debris and the debris velocity $v_{d}$ are two such parameters. As
we shall observe, the system makes several quick transitions from
quasi-periodic to chaotic states and vice versa when the control parameters
are varied. So, there are several non-chaotic states of the system
interspaced with numerous stages when the system undergoes torus breakdown
and becomes chaotic. These breakdowns can be very quick and dramatic
and do not follow a fixed pattern -- a hallmark of torus breakdown
\citep{greene,torus1,torus2}. One such transition phase is shown
in Fig.\ref{fig:Poincar-plots.} through the Poincar\'e plots.

The charge fluctuation of the external debris is modeled by periodically
switching the debris term on and off with a frequency $\nu$. Toward
this, the external debris term is expressed in a periodic top-hat
function as mentioned before
\begin{equation}
\rho_{{\rm ext}}(\tau)=\rho_{0}[1+\tanh\{\Delta\sin(\nu\tau)\}],
\end{equation}
where $\rho_{0}$ denotes of the strength of the charge density and
$\Delta$ controls the steepness of the top-hat function which is
typically $\sim8-10$. As our system is normalized, $\rho_{0}$ is
typically set to $\sim0.5$, which is about $50\%$ of the background
plasma density. The fluctuation frequency $\nu$ is set to $\sim1-5$
or about five ion-plasma periods. In Fig.\ref{fig:Poincar-plots.},
we have shown torus breakdown of the system where, in the top row,
the Poincar\'e map of the initial deformed torus (quasi-periodic)
is shown, which makes a final transition to a chaotic state on the
right when the breakdown is complete. The successive transition from
the initial state to the final state is depicted in the smaller panels
at the lower rows for shaded region marked in the bigger panels. The
whole transition occurs within the interval of the debris velocity
$v_{d}\in[1.596,1.6]$. The torus breakdown is further confirmed by
the power spectrum density $P$ of the oscillation, shown in the left
panel of Fig.\ref{fig:Power-spectrum-density}, where one can see
the numerous random frequencies dominated by some prominent frequencies.
The final chaotic state is also confirmed by the maximal Lyapunov
exponent $\lambda$ shown in the right panel of Fig.\ref{fig:Power-spectrum-density},
which settles down at about $\sim0.03$.

%%%%%%%%%%%%

{The reason for the \emph{quick} transition to a chaotic regime in the case of PIC simulation compared to the analytical model  is that the analytical model is a reduced fluid model, which does not have any kinetic effects and provides the dynamics for  only a single fluid element. Whereas the PIC model inherently has all the possible kinetic effects such as nonlinear Landau damping, phase-mixing, and particle trapping and nonlinearities making it intrinsically fast and nonlinear. Besides, it shows the cumulative dynamics of millions of particles with many directions for making some of the Lyapunov exponents grow fast.}

%%%%%%%%%%%%

\subsection{KAM tori analysis}

We note that the Kolmogorov--Arnold--Moser (KAM) theory \citep{kam}
describes how invariant tori exists in nearly integrable Hamiltonian
systems. In typical cases, these invariant tori break up through destruction
of homoclinic orbits (so-called homoclinic tangles) leading to chaos,
which is known as torus breakdown \citep{torus1,torus2}. However,
before we can qualify the nature of the chaos in our system as torus
breakdown (seems possible from the numerical solutions from the previous
section), we need some more clarifications in this regard.

\subsubsection{Preliminary analysis}

Let us write down our system represented by Eq.(\ref{eq:combined})
in a generalized form
\begin{equation}
\ddot{x}A(x)+\dot{x}^{2}B(x)+C_{0}(x)+\epsilon f(\tau)=0,\label{eq:x}
\end{equation}
where $x=v_{i}$ and the coefficient $C(x,\tau)$ is broken up into
an autonomous and a time-dependent part
\begin{eqnarray}
C_{0}(x) & = & \frac{\mu_{0}}{(v_{d}-x)}-e^{\phi(x)},\\
\epsilon f(\tau) & = & \rho_{{\rm ext}}(\tau).
\end{eqnarray}
Here, we have treated the external charged debris as a perturbation
with a smallness parameter $\epsilon$. The equivalent unperturbed
planar system is given by
\begin{equation}
\begin{array}{rclcl}
\dot{x} & = & y,\\
\dot{y} & = & F_{0}(x,y) & = & {\displaystyle \frac{-y^{2}B(x)-C_{0}(x)}{A(x)}}.
\end{array}
\end{equation}
The equilibrium of the above system implies that $y=C_{0}(x)=0$.
So, the linearized Jacobian of the vector field $(y,F_{0})$ at a
point $(x_{s},y_{s})\equiv(x_{s},0)$ in equilibrium is given by
\begin{equation}
J=\left(\begin{array}{cc}
0 & 1\\
J_{21} & 0
\end{array}\right),\quad J_{21}=-\frac{C_{0}'(x_{s})}{A(x_{s})},
\end{equation}
\begin{figure}[t]
\begin{centering}
\includegraphics[width=0.5\textwidth]{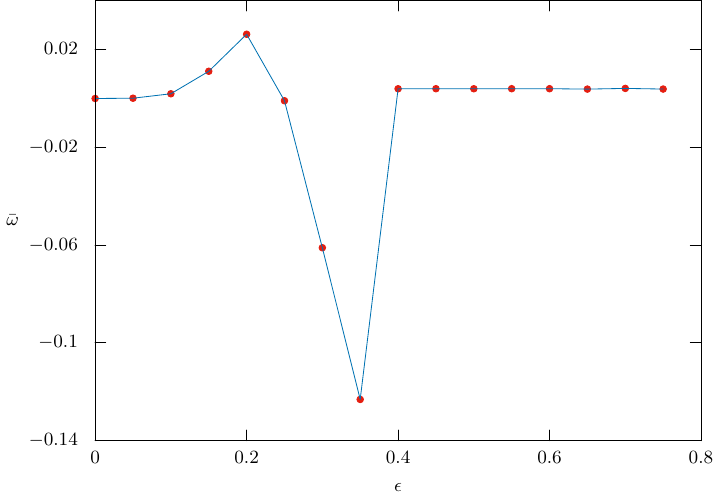}
\par\end{centering}
\caption{\protect\label{fig:The-frequency-map}The frequency map signifying
the transition to chaos through resonance.}
\end{figure}
\noindent\!\!\!\!\! with the eigenvalues $\varrho$ of the Jacobian as $\varrho=\pm\sqrt{J_{21}}.$
The equilibrium point $x_{s}$ can be determined from the condition
$C_{0}(x)=0$, and it comes out to be $x_{s}\simeq1.05935$, which
shows that $J_{21}<0$ and the existence of a saddle point and a homoclinic
orbit can be ruled out for the unperturbed system \citep{strogatz-book}.
Apparently Eq.(\ref{eq:combined}) is a highly nonlinear system and
as it contains a $\dot{v}_{i}^{2}$ term, it is not likely to be reducible
to a Hamiltonian system as well, even in the extended phase space
$(v_{i},\dot{v}_{i},\tau)$. This is also obvious from the fact that
the flow is `non volume-preserving' i.e.\ $\nabla\cdot\dot{\bm{x}}\neq0$,
where $\bm{x}=(x,y)$, which can be readily found out by observing
that
\begin{equation}
\nabla\cdot\dot{\bm{x}}\equiv\frac{\partial\dot{x}}{\partial x}+\frac{\partial\dot{y}}{\partial y}=-2y\frac{B(x)}{A(x)}\neq0.
\end{equation}
All these facts indicate that a typical breakdown of KAM tori is \emph{not}
possible in our system. However, strong signature of torus breakdown
suggests that torus breakdown may still occur, though not through
tangles but via shear and resonance \citep{kam1,kam2,kam3}. This
insight leads us to carry out an analysis through frequency and finite-time
Lyapunov exponent (FTLE) maps \citep{ftle1,ftle2} in the next subsection.
The above analysis also shows that the system does not preserve phase
space volume and dynamically acts like a dissipative system though
physically it does not have any dissipation.

\begin{figure}[t]
\begin{centering}
\includegraphics[width=0.51\textwidth]{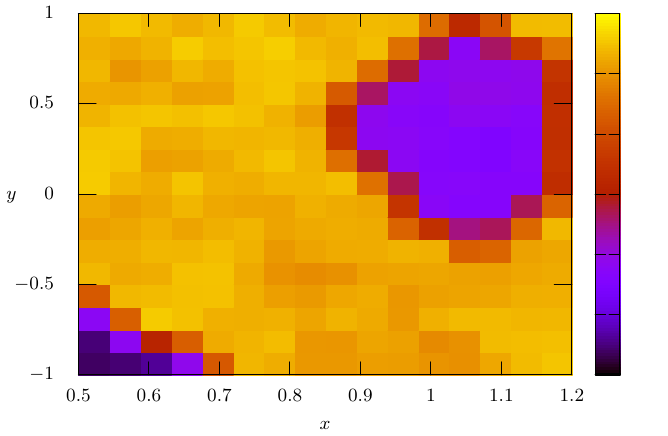}\hskip-12pt\includegraphics[width=0.51\textwidth]{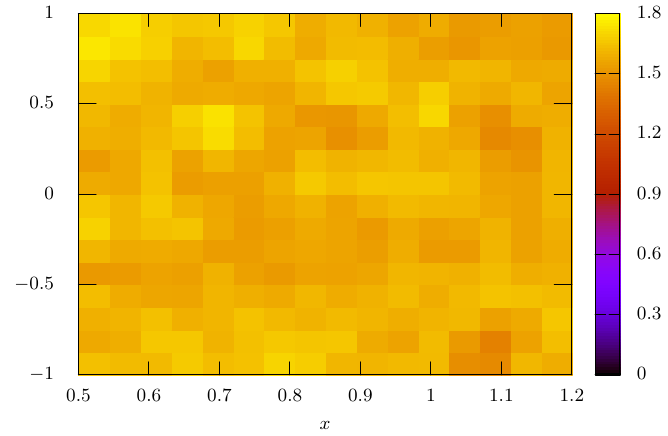}
\par\end{centering}
\caption{\protect\label{fig:The-FTLE-maps}The FTLE maps for positive forcing
(left) and negative forcing (right).}
\end{figure}

\subsubsection{Frequency and FTLE maps}

We now express our external charged debris term as a perturbation,
as mentioned before $\rho_{{\rm ext}}(\tau)=\epsilon f(\tau)$, where
$\epsilon$ is small but $\gtrless0$, depending on the nature of
charged perturbation. In order to compute the frequency map, we consider
the actual time evolution of $x(\tau)$, as obtained from the numerical
solution of Eq.(\ref{eq:x}) which is always real. From this, we construct
the analytic (or complex) signal $z(\tau)$ by incorporating the Hilbert
transform ${\cal H}[x(\tau)]$ of $x(\tau)$,
\begin{equation}
z(\tau)=x(\tau)+i{\cal H}[x(\tau)],
\end{equation}
This helps us compute the instantaneous frequency $\omega(\tau)$
\begin{equation}
\omega(\tau)=\frac{d}{d\tau}\theta(\tau),\quad\textrm{where}\,\theta(\tau)=\arg[z(\tau)].
\end{equation}
We however use the average instantaneous frequency $\bar{\omega}$,
averaged over a certain time $T$ to cancel out the transient fluctuations
\begin{equation}
\bar{\omega}=\frac{1}{T}\int_{T_{0}}^{T}\omega(\tau)\,d\tau.
\end{equation}
In Fig.\ref{fig:The-frequency-map}, we show a plot of $(\bar{\omega},\epsilon)$
with $\epsilon>0$ (positive charge perturbation) for a region of
torus breakdown, shown in Fig.\ref{fig:Poincar-plots.}. The figure
shows that in the region $\epsilon\lesssim0.2$ and $\gtrsim0.4$,
the instantaneous average frequency varies smoothly signifying quasi-periodic
motion. In the region $\epsilon\simeq0.25-0.35$, we have a sharp
transition as well as some fluctuations indicating the onset of chaotic
motion, which indicates a torus breakdown with possible resonance.

In order to compute the FTLE maps, we compute the Lyapunov exponent
by calculating the divergence of orbits over a finite time window
$T$, giving the FTLE \citep{ftle1,ftle2}
\begin{equation}
\lambda_{T}(\bm{x}_{0})=\frac{1}{T}\log\frac{\left\Vert \delta(T)\right\Vert }{\left\Vert \delta(0)\right\Vert },
\end{equation}
where $\delta(0)$ is the initial separation of two nearby trajectories
and $\delta(T)$ is the separation of the orbits at time $T$. We
note that the maximal Lyapunov exponent can be recovered through $\lambda_{{\rm max}}=\lim_{T\to\infty}\lambda_{T}(\bm{x}_{0})$.
The two FTLE maps -- one for positive charged perturbation and the
other for negative, are shown in Fig.\ref{fig:The-FTLE-maps}. As
we can see that for positive charge perturbation (positive forcing
with $\epsilon=+0.1$), there are dark basins with very low FTLEs
on either sides with surrounding yellow-orange colored high-FTLE region.
This sharp contrast is a definitive signature for large lobes of mixing
of tori, indicating a torus breakdown. On the other hand, negative
charge perturbation (negative forcing with $\epsilon=-0.1$) does
produce chaotic regions but as we can see from the right panel, the
map is more or less uniform. The FTLE maps further confirm our earlier
speculation that a positive charged perturbation produces very pronounced
fractal scattering indicating stronger chaos while a negative charge
perturbation of the same amplitude produces a milder chaos.

\begin{figure}[t]
\begin{centering}
\includegraphics[width=0.51\textwidth]{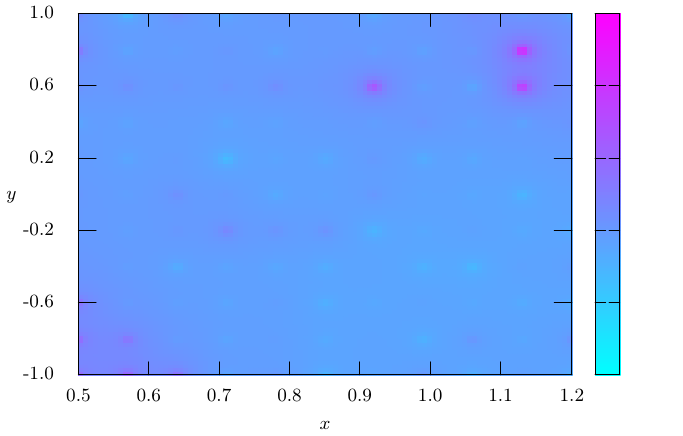}\hskip-12pt\includegraphics[width=0.51\textwidth]{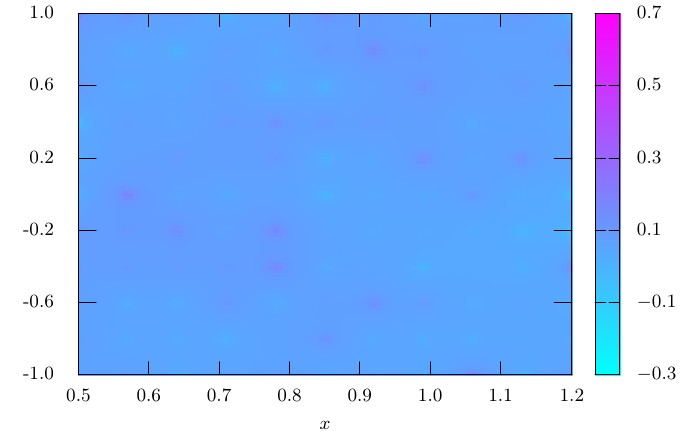}
\par\end{centering}
\caption{\protect\label{fig:The-FTLE-difference}The FTLE difference maps showing
effect of stochastic fluctuation for positive forcing (left) and negative
forcing (right).}
\end{figure}

\subsubsection{Perturbation-dependent chaos}

It is now amply clear from the numerical as well as theoretical analyses
that the chaotic behavior of the plasma system with external charged
debris is highly dependent on the nature of the charged debris (positive
and negative). Physically, we can perhaps understand it by rewriting
Eq.(\ref{eq:combined}) as
\begin{equation}
A(v_{i})\ddot{v}_{i}=-\dot{v}_{i}^{2}B(v_{i})-C(v_{i},\tau),
\end{equation}
which represents the total force of the system with $A(v_{i})$ acting
as a mass. As $C(v_{i},\tau)=C_{0}(v_{i})+\rho_{{\rm ext}}(\tau)$,
total forcing is larger when $\rho_{{\rm ext}}(\tau)>0$, while the
forcing is reduced when $\rho_{{\rm ext}}(\tau)<0$ with a possibility
of even cancelling the force rendering the system to a force-free
state. So, positive charge perturbation pushes the system away from
an equilibrium which results in a stronger chaotic state.

\subsubsection{Effect of stochastic fluctuations\protect\label{subsec:Effect-of-stochastic}}

In practice, we would generally expect a stochastic component to the
fluctuating charge density of the external debris, which can be modeled
as a noise to the $\rho_{{\rm ext}}(\tau)$
\begin{equation}
\rho_{{\rm ext}}(\tau)=\rho_{{\rm det}}(\tau)+\epsilon_{{\rm noise}}\xi(\tau),
\end{equation}
where $\epsilon_{{\rm noise}}$ denotes the strength of the noise
(or stochastic part) while the first term denotes the deterministic
part. In Fig.\ref{fig:The-FTLE-difference}, we show the FTLE difference
maps for two cases $\epsilon=\pm0.1$ where the difference between
the usual FTLE and one with stochastic component is shown. We can
see from the figure that in both cases $(\epsilon=\pm0.1)$, the noise
tends to destabilize weakly regular (non-chaotic or marginally chaotic)
areas, whereas highly chaotic zones are least affected. So, stochastic
forcing primarily affects the marginal or near-integrable areas and
overall chaotic property of the system remains broadly same, a result
which is confirmed from our PIC simulation.

\begin{figure}[t]
\begin{centering}
\includegraphics[width=0.51\textwidth]{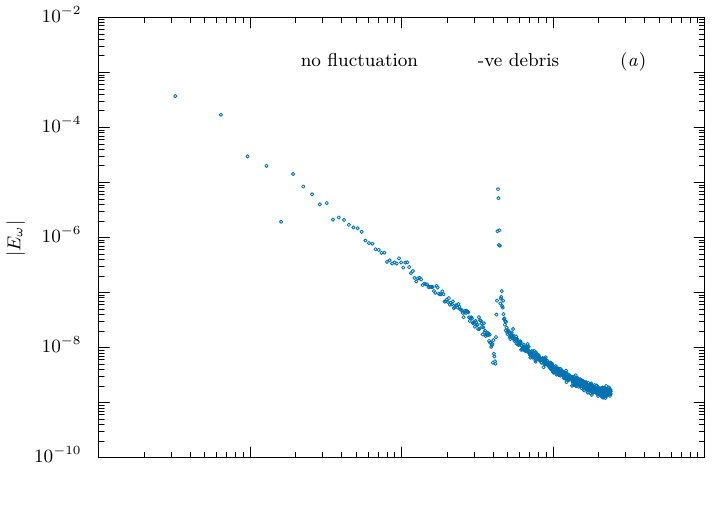}\includegraphics[width=0.51\textwidth]{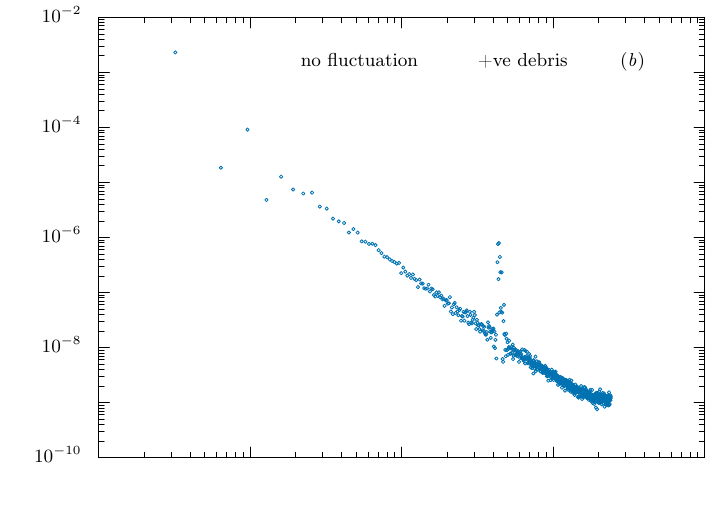}\\
~\vskip-36pt~\\
\includegraphics[width=0.51\textwidth]{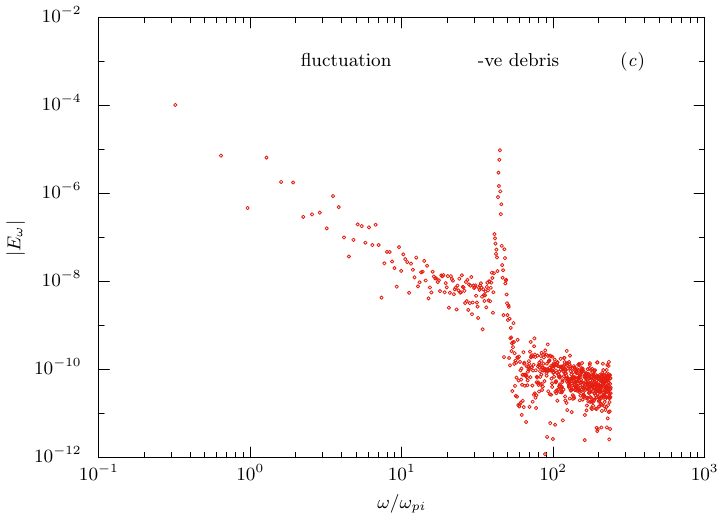}\includegraphics[width=0.51\textwidth]{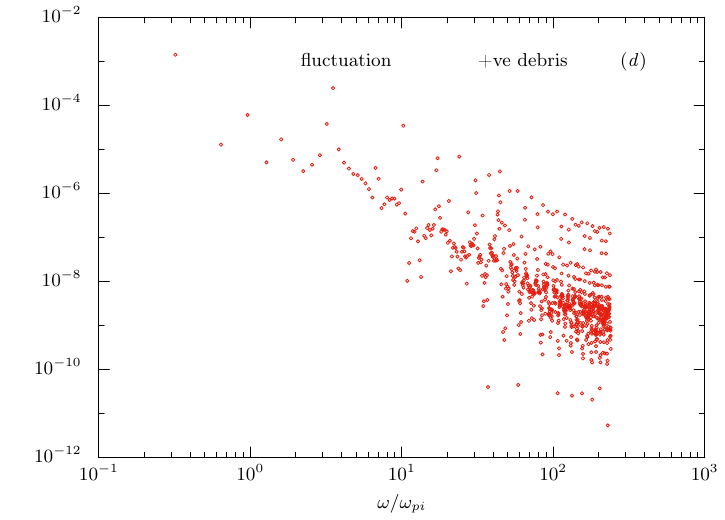}
\par\end{centering}
\caption{\protect\label{fig:The-ion-energy}The ion energy spectrum $|E_{\omega}|\sim\left(v_{i}^{2}\right)$
for both negatively and positively charged debris with and without
fluctuation. The upper blue colored panels (\emph{a}) and (\emph{b})
are for negatively and positively charged debris, respectively without
any periodic charge fluctuation of the debris. The lower red colored
panels (\emph{c}) and (\emph{d}) are for negatively and positively
charged debris with periodic charge fluctuation of the debris.}
\end{figure}

\section{Spectral signature of the phase space dynamics}

In this section, we provide an analysis on the basis of spectral signatures
of the chaotic phase space dynamics from the PIC simulation data.
The analysis broadly confirms our earlier observations given in Sections
\ref{sec:Simulation-results} and \ref{sec:Theoretical-analysis}.
In Fig.\ref{fig:The-ion-energy}, we have shown the ion energy spectrum
$|E_{\omega}|\sim\left(v_{i}^{2}\right)$ for both negatively and
positively charged debris with and without debris charge fluctuation.
The first observation is that the energy spectra becomes flattened
at higher frequencies which points to the energy redistribution among
the waves and particles due to linear and NLLD \citep{nlandau,landau2,landau1}.
We note that the system has no physical dissipation which would have
steepened the spectra at the right ends indicating loss of energy.
However, in absence of any physical dissipation, the Landau damping
merely redistributes the energy keeping the total energy content of
the system to remain constant, which causes this flattening.

The other important observation is the broadening of the spectrum
in presence of periodic charge fluctuation of the debris. This broadening
indicates that the coherent cascading of energy is no longer possible
when there is a periodic charge fluctuation of the debris. This also
points toward the limitation imposed by chaotic oscillations on coherent
transfer of energy from wave to particle and vice versa. The higher
broadening of the spectrum for fluctuation of positively charged debris
{[}panel (\emph{d}){]} indicates more loss of coherence due to a stronger
chaos \citep{broad2,broad1}. These energy spectra also suggest that
the nonlinear structures produced due to negatively charged debris
are dynamically more robust and are less prone to any charge fluctuation
of the debris.

The spike at $\omega\sim43\omega_{pi}$ in the spectra is due to the
electron plasma frequency which basically shows that the primary driven
mechanism of the ion oscillation is due to the electron response.

\section{Conclusions}

In this work, we have looked into the dynamics governing the interplay
between external charged debris and a flowing $e$-$i$ plasma with
a particular emphasis on the role of the periodic fluctuation of debris
charge. Our analysis is focused on how these fluctuations, introduced
as an external forcing in our model, affect the plasma response. We
note that periodic charge fluctuation on external debris can occur
naturally due to a feedback response to the IAW induced by the presence
of the external debris. Our analysis shows emergence of complex nonlinear
phenomena including the emergence of chaos and the manifestation of
NLLD.

We took a dual simulation approach to unravel these dynamics using
a kinetic particle-in-cell (PIC) simulation and flux-corrected transport-based
(FCT) fluid simulation. Our findings reveal that the charge polarity
of the debris plays a crucial role in determining the system\textquoteright s
behaviour. The charge fluctuation of the debris is implemented through
a periodic external forcing model in both the simulations with the
charge polarity fixed as either positive or negative throughout the
simulations. Some key findings of our simulations are:
\begin{itemize}
\item The debris charge fluctuation acts as the trigger for chaotic oscillations,
regardless of the charge polarity and the relative velocity between
the plasma and debris.
\item For a positively charged debris, we have observed a much stronger
chaotic response confined at the debris site, with the maximal Lyapunov
exponent increasing with the debris velocity. In contrast, negatively
charged debris creates a more uniform weak chaos pattern across the
domain.
\item When the frequency of debris velocity approaches IA speed $v_{d}\sim c_{i}$,
both PIC and FCT simulations detect damping of the nonlinear wave,
which can be due to both nonlinear wave-particle resonance (NLLD)
as well as nonlinear wave-wave interactions. However, while FCT simulation
can only show wave-wave interactions, PIC simulation can show both
the effects. The existence of NLLD is validated from the flattening
of the ion velocity distribution.
\item The added stochastic component to the regular periodic fluctuation
does not affect the chaotic ion-acoustic oscillations significantly.
\end{itemize}
Theoretically, we have shown that the emergence of chaos due to the
periodic charge fluctuation of the debris is through torus breakdown,
though the breakdown happens via shear and resonance \citep{kam1,kam2,kam3},
not through the usual process of destruction of homoclinic orbits.
The primary results of the theoretical analysis are:
\begin{itemize}
\item The periodic fluctuation of debris charge in a flowing plasma leads
to chaotic ion-acoustic oscillations and the chaos occurs through
a phase of torus breakdown. This is substantiated by a KAM tori analysis.
\item The reason why a periodically fluctuating negative debris charge induces
a much weaker chaos than positive debris charge is due to the fluctuation
forcing the system away from a dynamically stable configuration when
the debris charge is positive.
\item Through the FTLE maps, the theoretical analysis has also been able
to show why the stochastic component affect the chaos only negligibly.
\end{itemize}
At the end, we have carried out an energy spectrum analysis of the
chaotic phase space, which further confirms the dynamic stability
of the nonlinear structures induced by negatively charged debris.
Our work adds a new aspect to the science of the detection of LEO
space debris. While existing methods leverage on plasma oscillations
induced by the charged debris, the different chaotic signatures for
positive and negatively charged debris as we have observed here add
another layer to the detection scenario which can make the management
of space debris even more robust.

\section*{Acknowledgement}
BJ would like to acknowledge Gauhati University for research scholarship grant GU/UGC/GURF/2023/712. HS would like to thank CSIR-HRDG, New Delhi, India for Senior Research Fellowship research grant 09/059(0074)/2021-EMR-I.

The authors would also like to thank the anonymous referees as their critical comments and suggestion have greatly improved the manuscript.

\appendix
\section{Collisionality and kinetic simulation}
{It can be argued about the validity of the kinetic simulation in the light of the regime of collisionality. In this appendix, we establish the validity of the kinetic modeling taking consideration of the $e$-$i$ collision by evaluating the Knudsen number.
We can derive the Knudsen number ($\textrm{Kn}$) from the first principle
in plasma considering the fact that the dominant collisions are between
electrons and ions (this causes the highest momentum transfer). So,
by definition, we have
\begin{equation}
\textrm{Kn}=\frac{\lambda_{\textrm{mean}}}{L},
\end{equation}
where $\lambda_{\textrm{mean}}$ is the mean free path for $e$-$i$
collisions and $L$ is the characteristic length of the system. As
compared to the ions, electrons are highly mobile (in our case as
well), we can assume that
\begin{equation}
\lambda_{\textrm{mean}}\simeq\frac{v_{\textrm{th}e}}{\nu_{ei}},
\end{equation}
where the electron thermal velocity, $v_{\textrm{th}e}=\sqrt{T_{e}/m_{e}}$ and electron-ion collision frequency, $\nu_{ei}$ for Coulomb cross section can be approximated as \citep{chen-1974}
\begin{equation}
\nu_{ei}\sim\frac{ne^{4}}{16\pi\epsilon_{0}^{2}m_{e}^{2}v_{\textrm{th}e}^{3}},
\end{equation}
where $n$ is the average plasma density. Our simulation box is of length $L\sim140\lambda_{D}\sim1\,\textrm{cm}$.
Plugging all the numbers for our simulation parameters, we arrive
at the value for the Knudsen number as
\begin{equation}
\textrm{Kn}\sim15\gg1,
\end{equation}
which shows that the regime is almost collision-free and the use of
kinetic theory is quite justified (through the PIC model).

It can further be argued about the use of the fluid model  (Section 2.4) when $\textrm{Kn}\gg1$. We note that the fluid equations used in
this work (or in plasma physics contexts as well) are weighted average
of the Boltzmann-Vlasov equation  rather than
the collisional Boltzmann equation (using Chapman-Enskog expansion).
Naturally these fluid equations are valid \emph{even} in the limit of zero
collisions, which is due to the fact that the plasma dynamics is governed
by long range Coulomb force rather than Newtonian collisional dynamics
(like a hydrodynamic gas).}

%\bibliographystyle{elsarticle-num}
%\bibliography{ref}

%\end{linenumbers}

%bibliography from .bbl file

\end{document}